\documentclass[12pt]{article}

\usepackage{amsmath}
\usepackage{euscript}

\usepackage{pstricks,epsfig,amssymb,axodraw}
\usepackage{cite}  

\def\Order#1{${\cal O}(#1)$}

\def\Order#1{${\cal O}(#1)$}

\def\st{\hbox{}} 

\newcommand{\Meu}{\EuScript{M}}

\newcommand{\Bmf}{\mathfrak{B}}

\newcommand{\sfac}{\mathfrak{s}}

\begin{document}

\begin{titlepage}
\thispagestyle{empty}

\begin{flushright}
{\bf  
HNINP-V-04-04\\
} 
\end{flushright}
 
\noindent
\vspace*{5mm}

\begin{center}
{
     \Large\bf Gauge invariance, infrared/collinear singularities and \\
tree level matrix element for  $e^+ e^- \to \nu_e \bar \nu_e \gamma \gamma$.}
\end{center}


\vspace*{2mm} 

\begin{center}
 
  {\bf Z. W\c{a}s }

  \vspace{5mm}
  {\em  Institute of Nuclear Physics PAS,
           ul. Radzikowskiego 152, PL-31-342 Cracow, Poland}\\

\end{center}
\vfill

\begin{abstract}
One of the necessary steps in constructing high precision option of 
KKMC was to install the double bremsstrahlung matrix element for the 
process $e^+ e^- \to \nu_e \bar \nu_e $ into the scheme
of Coherent Exclusive Exponentiation. The process is also interesting
because of gauge cancellation of contributions for photon emission 
 from incoming fermion lines  and  $t$-channel $W$. The QED U(1)  gauge properties
require terms  of the triple and quatric gauge couplings to be taken into
considerations as well. Thanks   
to expansion starting from the approximation of contact interaction,
good example to study the internal structure of the amplitude is available. 

In the developed scheme,  natural
separation of the complete amplitude into gauge invariant parts is 
straightforward. 
Each part has well defined physical
interpretation, which after partial integration
over phase space provides terms:  infrared singular, 
leading log, next-to-leading-log, etc.
Contributions related to triple and quatric gauge coupling of  $W$
(extracted with the help of expansion around contact $W$-interaction),  
have been ordered as well. 
The separation is also helpful, to define extrapolation/reduction procedure 
of CEEX exponentiation for the $\nu_e$ channel.

\end{abstract}  
\vspace*{1mm}

\begin{center}
\end{center}

\vspace*{1mm}
\vfill
\begin{flushleft}
{\bf HNINP-V-04-04 \\
  May 2004 }
\end{flushleft}

\vspace*{1mm}
\bigskip
\footnoterule
\noindent
{\footnotesize \noindent
Work supported in part by the European Union 5-th Framework under contract HPRN-CT-2000-00149.
Partly supported by
Marie Curie Host Fellowship for the Transfer of Knowledge
Contract No. MTKD-CT-2004-510126.
}
\end{titlepage}

\normalsize

\section{Introduction}

Higher order radiative corrections
are usually necessary
to obtain from Standard Model high precision results for phenomenologically important quantities. 
The techniques of direct calculations, lead to expressions, of often
hundreds, thousands or even million's of terms.  These are difficult 
 to control analytically  and/or  numerically. This is worrisome, because to 
obtain phenomenologically sound results third order effects 
are mandatory, see e.g.  \cite{Jadachtransp}. This is  clearly outside the reach of 
presently available methods of direct perturbative calculations. 
There is no doubt, that resummation is  necessary of at least some 
contributions from orders higher than the second one.

In case of electroweak processes at LEP techniques based on exclusive 
exponentiation of  QED effects turned out to be powerful and enabled 
high precision predictions 
for a wide game of processes, such as Bhabha scattering, 
production of heavy bosons $W$ or $Z$ and lepton pairs.  The underlying method,
originating from pioneering work  of Yennie Frautschi and Suura \cite{yfs:1961} 
turned out to be realizable \cite{yfs1:1988,yfs2:1990,Jadach:1994yv,Jadach:1997is,Jadach:1999vf} in practice,
 thanks to acumulated experience and ever increasing computer power.

One of the necessary elements in  approach based on exponentiation 
is rigorous study of matrix elements obtained from perturbative calculation. 
In fact it is not enough to calculate predictions at 
as high order of perturbation expansion as possible, but also to carefully separate 
results into infrared singular and remaining finite parts. 
Thanks to the properties of QED, results of explicit perturbative calculations are not necessary
to obtain singular and leading terms of every order.
These leading parts of amplitudes can be
combined with the phase space into the module of the low level Monte Carlo
generator or in general into multi-dimensional distribution which can be understood
as lowest order of improved perturbative expansion. Later, finite parts of the  matrix elements can be added 
order by order.  In case of Monte Carlo algorithms it can be done with the help of the correcting 
weight, which
can be shown to be positive and bound from above.
Details of such a rigorous scheme  can be found  
in refs. \cite{Jadach:1999vf,Jadach:2000ir}. It improves significantly the convergence
of perturbative expansion; final states with  arbitrary
number of photons are present already at the lowest level of expansion.  
 This allows for predictions with  realistic 
experimental cut-offs included.  
The solution based on separation performed at the amplitude
level, is specially useful. It opens
the way for easy implementation of all sort of interferences, 
also convergence of expansion is particularly fast in this case. 
The underlying exponentiation scheme is called Coherent Exclusive Exponentiation (CEEX).

The following point is of practical importance. In case of exponentiation,
already at the lowest order of expansion, configurations with multiple 
real photons are present.
That is why, it may happen, that for  particular event, there is more 
explicit photons in final state, than in expression directly available 
from standard
perturbative expansion.  
Reduction/extrapolation methods are then necessary.
We will not elaborate much on theoretical aspects of this point here.
However let us stress that
if sufficiently high order of perturbation expansion
is available, dependence on the choice of reduction procedure
or extrapolation is dropping out. Particularly bad choice may
nonetheless degrade convergence of expansion. 
It is thus of the importance to provide results of perturbative calculations 
in a form as convenient for extrapolation procedure as possible. 
Comparisons of amplitudes calculated at different orders of perturbative expansion can provide a useful hint.

In the present paper, we heavily rely on  ref.~\cite{Jadach:1998wp}.
 We will assume certain
level of  familiarity  of the reader with that reference. Also, let us note
that spin amplitudes for the
process $e^+ e^- \to \nu_e \bar \nu_e \gamma \gamma$ are well defined
within the Standard Model and known since long, see  e.g. \cite{Berends:1988zz}. We 
could profit in our work from the ready to use computer codes such as 
\cite{Kurihara:1999vc} available for numerical cross checks of our results.

Our paper is organized as follows. Section 2 is devoted to the case of single 
bremsstra\-hlung  $e^+ e^- \to \nu_e \bar \nu_e \gamma $, some  basic elements
 of spin amplitude techniques, useful in the more complex case of double bremsstrahlung
are presented there as well.
In section 3, we provide the main results, in particular  we explicitly
identify gauge invariant parts of the amplitudes. We stress
 points which will be useful for  extrapolation schemes used in 
CEEX exponentiation as well.
We keep our discussion, having in mind future
applications in  spin-amplitude automated programs. 
  In section 4 we discuss issues related to extrapolation procedure in more detail.
Finally, Section 5 closes the paper.

\section{Amplitude for one real photon and notation}
\begin{figure}[h]
\centering
\setlength{\unitlength}{1mm}
\begin{picture}(180,85)
\put( -75, -35){\makebox(0,0)[lb]
{\epsfig{file=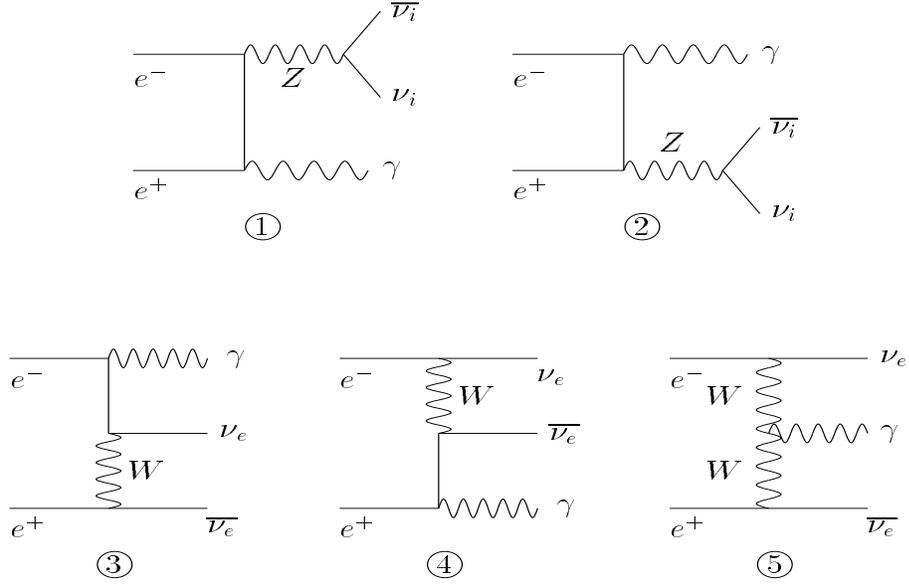,width=280mm,height=350mm}}}
\end{picture} 
 \caption{{\em The Feynman diagrams for
          $e^+ e^- \rightarrow \bar \nu_e \nu_e \gamma$.}}
 \label{fig:bremI}
\end{figure}

Let us start with  the well-known and straightforward to calculate by any method 
\Order{\alpha} spin amplitude for
the  $e^+e^- \to \nu_e \bar\nu_e \gamma$ single-photon brems\-strahlung process, 
see fig~\ref{fig:bremI}. 
We will recall it nonetheless here, to define framework  for our 
discussion.  We will use
conventions of  refs.~\cite{Jadach:2000ir,Bardin:2001vt}. Let us recall here
only the most important notations.
The four momenta $p_a$, $p_b$, $p_c$, $p_d$, $k_1$ denote respectively
momenta of incoming electron, positron, outcoming neutrino, antineutrino and finally photon.
The indices for the spin states for the fermions are denoted
 respectively as $\lambda_a$, $\lambda_b$, 
$\lambda_c$, $\lambda_d$ and for photon $\sigma_1$. The photon polarization vector
is denoted as $\epsilon_{\sigma_1}$. The gauge transformation in our case  reduces
to the replacement  $\epsilon_{\sigma_1} \to \epsilon_{\sigma_1} + {x}\  k_1 $
(with arbitrary coefficient $x$), there will be no external line bosons, and incoming
fermions lead to the trivial phases only.
With these notations the first-order matrix element%
\footnote{ $\Meu_{1\{I\}}\left( \st^{p}_{\lambda} \st^{k_1}_{\sigma_1} \right)$ 
The subscripts $1$ and $\{I\}$ denote respectively, that the amplitudes are of the first order and are included 
as part of the initial state bremsstrahlung. This spurious notation is however convenient for the reader interested
in ref.~\cite{Jadach:1998wp}.}
obtained  from the Feynman diagrams  
depicted in fig.~\ref{fig:bremI}, can be written in a rather straightforward way:
\begin{equation}
  \label{isr-feynman}
  \begin{split}
    \Meu_{1\{I\}}\left( \st^{p}_{\lambda} \st^{k_1}_{\sigma_1} \right)=
   & eQ_e \;
    \bar{v}(p_b,\lambda_b)\; \mathbf{M}^{bd}_{\{I\}}\;
         {\not\!{p_a}+m-{\not\!k_1} \over -2k_1p_a} \not\!{\epsilon}^\star_{\sigma_1}(k_1)\;
    u(p_a,\lambda_a)\\
   +&eQ_e \;
    \bar{v}(p_b,\lambda_b)
         \not\!{\epsilon}^\star_{\sigma_1}(k_1)\; {-{\not\!p_b}+m+\not\!{k_1} \over -2k_1p_b} \;
         \mathbf{M}^{ac}_{\{I\}}\;
    u(p_a,\lambda_a) \\
    +&e \;
    \bar{v}(p_b,\lambda_b)
        \; \;
         \mathbf{M}^{bd,ac}_{\{I\}}\;
    u(p_a,\lambda_a) \;\;\;
{ {\epsilon}^\star_{\sigma_1}(k_1) \cdot (p_c-p_a+p_b-p_d)
 \over (t_a -M_W^2) (t_b -M_W^2)} \\      \;
   +&e \; {\bar{v}(p_b,\lambda_b)  g_{\lambda_b,\lambda_d}^{We\nu}
         \not\!{\epsilon}^\star_{\sigma_1}(k_1)\;  \;
    v(p_d,\lambda_d)
    \bar{u}(p_c,\lambda_c)  g_{\lambda_c,\lambda_a}^{We\nu}
         \not\!k_1\;  \;
    u(p_a,\lambda_a) \over (t_a -M_W^2) (t_b -M_W^2)}\\
   -&e \; { \bar{v}(p_b,\lambda_b)  g_{\lambda_b,\lambda_d}^{We\nu}
         \not\!k_1\;  \;
    v(p_d,\lambda_d)
    \bar{u}(p_c,\lambda_c) g_{\lambda_c,\lambda_a}^{We\nu}
         \not\!{\epsilon}^\star_{\sigma_1}(k_1)\;  \;
    u(p_a,\lambda_a) \over (t_a -M_W^2) (t_b -M_W^2)},
  \end{split}
\end{equation}
or, equivalently:
\begin{equation}
  \label{isr-feynman1}
  \begin{split}
   &\Meu_{1\{I\}}\left( \st^{p}_{\lambda} \st^{k_1}_{\sigma_1} \right)=
    {\cal M}^0+{\cal M}^1+{\cal M}^2+{\cal M}^3
   \\&
    {\cal M}^0=
   eQ_e \;
    \bar{v}(p_b,\lambda_b)\; \mathbf{M}^{bd}_{\{I\}}\;
         {\not\!{p_a}+m-{\not\!k_1} \over -2k_1p_a} \not\!{\epsilon}^\star_{\sigma_1}(k_1)\;
    u(p_a,\lambda_a)
    \\&
   +eQ_e \;
    \bar{v}(p_b,\lambda_b)
         \not\!{\epsilon}^\star_{\sigma_1}(k_1)\; {-{\not\!p_b}+m+\not\!{k_1} \over -2k_1p_b} \;
         \mathbf{M}^{ac}_{\{I\}}\;
    u(p_a,\lambda_a) 
    \\&
       {\cal M}^{1}={\cal M}^{1'}+{\cal M}^{1''}   \\&
       {\cal M}^{1'}=+e \;
    \bar{v}(p_b,\lambda_b)
        \; \;
         \mathbf{M}^{bd,ac}_{\{I\}}\;
    u(p_a,\lambda_a) {\epsilon}^\star_{\sigma_1}(k_1) \cdot (p_c-p_a) {1 \over {t_a -M_W^2}}{1 \over {t_b -M_W^2}},
    \\&
       {\cal M}^{1''}=+e \;
    \bar{v}(p_b,\lambda_b)
        \; \;
         \mathbf{M}^{bd,ac}_{\{I\}}\;
    u(p_a,\lambda_a) {\epsilon}^\star_{\sigma_1}(k_1) \cdot (p_b-p_d) {1 \over {t_a -M_W^2}}{1 \over {t_b -M_W^2}},
    \\&
    {\cal M}^2=   +e \; \bar{v}(p_b,\lambda_b)  g_{\lambda_b,\lambda_d}^{We\nu}
         \not\!{\epsilon}^\star_{\sigma_1}(k_1)\;  \;
    v(p_d,\lambda_d)
    \bar{u}(p_c,\lambda_c)  g_{\lambda_c,\lambda_a}^{We\nu}
         \not\!k_1\;  \;
    u(p_a,\lambda_a) {1 \over {t_a -M_W^2}}{1 \over {t_b -M_W^2}}
    \\&
     {\cal M}^3=  -e \; \bar{v}(p_b,\lambda_b)  g_{\lambda_b,\lambda_d}^{We\nu}
         \not\!k_1\;  \;
    v(p_d,\lambda_d)
    \bar{u}(p_c,\lambda_c) g_{\lambda_c,\lambda_a}^{We\nu}
         \not\!{\epsilon}^\star_{\sigma_1}(k_1)\;  \;
    u(p_a,\lambda_a) {1 \over {t_a -M_W^2}}{1 \over {t_b -M_W^2}},
  \end{split}
\end{equation}
where, the part of the amplitude, consisting of bosonic couplings
($g_\lambda^{Z,f}$ denote coupling constant of  $Z$ with fermion $f$ and
 handedness $\lambda$, in electric charge units), final state fermion spinors and boson propagators reads
as 
\begin{equation}
 \mathbf{M}^{xy}_{\{I\}} =ie^2  ({\cal R}_Z+{\cal R}_W) =ie^2
                              \sum_{B = W,Z} \Pi^{\mu\nu}_B(X)\; 
                              G^{B}_{e,\mu}\; (G^{B}_{f,\nu})_{[cd]}
\end{equation}
with
\begin{equation}
  \begin{split}
&
G^{B}_{e,\mu}=\gamma_\mu \sum_{\lambda=\pm} \frac{1}{2}(1+\lambda \gamma_5) g_\lambda^{B,e}\\&
(G^{B}_{f,\nu})_{[cd]}=\bar u(p_c,\lambda_c) G^{B}_{f,\nu} v(p_d,\lambda_d)\\&
\Pi^{\mu\nu}_{B=Z}(X)=\frac{g^{\mu\nu}}{X^2-M_Z^2+i \Gamma_Z X^2/M_Z}\\&
\Pi^{\mu\nu}_{B=W}(X)=\frac{g^{\mu\nu}}{t-M_W^2}.
  \end{split}
\end{equation}

The final-state spinors 
are explicitly included, and Fierz transformation is applied for the
part of $W$ exchange. The $W$ coupling constant reads

\begin{equation}
 g_{\lambda_c,\lambda_a}^{We\nu} = {1 \over \sqrt{2} \sin\theta_W} 
\delta^{\lambda_c}_{\lambda_a} \delta^{\lambda_c}_+.
\end{equation}
Only for the $W$ contribution, the superscripts $xy$ in $ \mathbf{M}_{\{I\}}$ have the meaning,
they define the momentum
transfer in the $W$ propagator $\Pi^{\mu\nu}_W(X)$: for $xy=ac$
the transfer 
\footnote{ Transfers can be expressed also as
$t_a=(p_b-k_1 -p_d)^2$ and $t_b=(p_a -k_1-p_c)^2$,
this make difference if extrapolation procedures are used 
 for the configurations off  mass shell where $p_a+p_b \ne p_c+p_d+k_1$, otherwise 
${\cal M}^{1'}={\cal M}^{1''} $ of course.}
is $t_a=(p_a -p_c)^2$, for $bd$ it is  $t_b=(p_b -p_d)^2$. 
If  both are explicitly marked, then the expression
\begin{equation}
  \mathbf{M}^{bd,ac}_{\{I\}} =ie^2 
         G^{W}_{e,\mu}\; (G^{W,\; \mu}_{\nu})_{[cd]}
\end{equation}
is used. For that parts of formula  (\ref{isr-feynman1})
 $W$ propagators are explicitly given. The notations  ${\cal R}_Z$
and  ${\cal R}_W$ will be used later, see respectively  formula
(\ref{cdotsZ}) and (\ref{cdotsW}).

Let us start now  to rewrite  expression (\ref{isr-feynman1}). 
It is straightforward to notice
that the first term ${\cal M}^0$  can be split into soft 
IR parts proportional to $(\not\!{p} \pm m)$
and non-IR parts proportional to ${\not\!k_1}$. The  non-IR parts are 
individually gauge invariant by construction. The soft part
of ${\cal M}^0$, with $Z$ couplings
only, is gauge invariant as well.
  
Employing the completeness relations of eq.~(A14) from ref.\cite{Jadach:2000ir}
 we obtain the different form of  (\ref{isr-feynman1}):
\begin{equation}
  \begin{split}
    \Meu_{1\{I\}}
    \left( \st^{p}_{\lambda} \st^{k_1}_{\sigma_1}
    \right)=
   &-{eQ_e\over 2k_1p_a}\; \sum_{\rho_a}
     \Bmf\left[ \st^{p_b}_{\lambda_b}  \st^{p_a}_{\rho_a}\right]\st_{[cd]}
          U\left[ \st^{p_a}_{\rho_a}  \st^{k_1}_{\sigma_1}  \st^{p_a}_{\lambda_a} \right]
    +{eQ_e\over 2k_1p_b}\; \sum_{\rho_b}
           V\left[ \st^{p_b}_{\lambda_b} \st^{k_1}_{\sigma_1} \st^{p_b}_{\rho_b}  \right]
      \Bmf\left[ \st^{p_b}_{\rho_b}  \st^{p_a}_{\lambda_a} \right]\st_{[cd]}\\
   +&{eQ_e\over 2k_1p_a}\; \sum_\rho
     \Bmf\left[\st^{p_b}_{\lambda_b}  \st^{k_1}_{\rho}  \right]\st_{[cd]}
          U\left[\st^{k_1}_{\rho}        \st^{k_1}_{\sigma_1}  \st^{p_a}_{\lambda_a} \right]
    -{eQ_e\over 2k_1p_b}\; \sum_\rho
           V\left[ \st^{p_b}_{\lambda_b} \st^{k_1}_{\sigma_1}  \st^{k_1}_{\rho}      \right]
      \Bmf\left[ \st^{k_1}_{\rho}  \st^{p_a}_{\lambda_a}\right]\st_{[cd]}\\
+ &{\cal M}^{1'}+{\cal M}^{1''}+{\cal M}^2+{\cal M}^3.
  \end{split}
\end{equation}
The terms ${\cal M}^{1'}$ to ${\cal M}^3$ correspond to the
last three lines%
\footnote{%
  The term ${\cal M}^1+{\cal M}^2+{\cal M}^3$
  originates from the $WW\gamma$ vertex
  \begin{displaymath}
    -ie \bigl[  
    g_{\mu\nu}(p-q)_\rho +g_{\nu\rho}(q-r)_\mu +g_{\mu\rho}(r-p)_\nu 
    \bigr]
  \end{displaymath}
  where all momenta are outcoming, and
  indices on outgoing lines are paired with momenta as
  $p^\mu$, $q^\nu$ $r^\rho$;
  ${\cal M}^1$ originates from the term where $g^{\mu\nu}$ 
  connects the $e^-$--$\nu_e$, $e^+$--$\bar \nu_e$ fermion lines. 
  }
of eq.~(\ref{isr-feynman}). 
These contributions are also IR-finite. 
In the next step let us remove
the sum in the first two terms  thanks to 
 the diagonality of $U$ and $V$  ( ref.\cite{Jadach:2000ir}). 
The matrices $\Bmf$ are also defined in this reference. We obtain
\begin{equation}
  \label{first-order-isr}
  \begin{split}
   \Meu_{1\{I\}}
      \left( \st^{p}_{\lambda} \st^{k_1}_{\sigma_1} \right)
   =& \sfac^{\{I\}}_{\sigma_1}(k_1) 
  \hat  \Bmf\left[\st^{p}_{\lambda} \right]
+ \bigl(  r_{\{I\}}^{B'} + {\cal M}^{1'} \bigr)
+ \bigl(  r_{\{I\}}^{B'} + {\cal M}^{1''} \bigr)
   \\
   &+    
r_{\{I\}}^{A'}  +r_{\{I\}}^{A''}  + \bigl({\cal M}^2+{\cal M}^3\bigr)\\
   r_{\{I\}}^{B'} \left( \st^{p}_{\lambda} \st^{k_1}_{\sigma_1} \right)= 
   &-{eQ_e\over 2k_1p_a}\; \sum_\rho
     \bar{\Bmf}\left[ \st^{p_b}_{\lambda_b}  \st^{p_a}_{\rho_a}\right]\st_{[cd]}
          U\left[ \st^{p_a}_{\rho_a}  \st^{k_1}_{\sigma_1}  \st^{p_a}_{\lambda_a} \right]
\\
   r_{\{I\}}^{B''} \left( \st^{p}_{\lambda} \st^{k_1}_{\sigma_1} \right)= 
   &    +{eQ_e\over 2k_1p_b}\; \sum_\rho
           V\left[ \st^{p_b}_{\lambda_b} \st^{k_1}_{\sigma_1} \st^{p_b}_{\rho_b}  \right]
      \bar{\Bmf}\left[ \st^{p_b}_{\rho_b}  \st^{p_a}_{\lambda_a} \right]\st_{[cd]}\\
   r_{\{I\}}^{A'} \left( \st^{p}_{\lambda} \st^{k_1}_{\sigma_1} \right)=&
    +{eQ_e\over 2k_1p_a}\; \sum_\rho
     \Bmf\left[ \st^{p_b}_{\lambda_b} \st^{k_1}_{\rho}   \right]\st_{[cd]}
        U\left[   \st^{k_1}_{\rho} \st^{k_1}_{\sigma_1}  \st^{p_a}_{\lambda_a}   \right],\\
   r_{\{I\}}^{A''} \left( \st^{p}_{\lambda} \st^{k_1}_{\sigma_1} \right)=&
    -{eQ_e\over 2k_1p_b}\; \sum_\rho
        V\left[   \st^{p_b}_{\lambda_b} \st^{k_1}_{\sigma_1} \st^{k_1}_{\rho}   \right]
     \Bmf\left[ \st^{k_1}_{\rho}    \st^{p_a}_{\lambda_a} \right]\st_{[cd]},\\
   \sfac^{\{I\}}_{\sigma_1}(k_1) = &
     -eQ_e{b_{\sigma_1}(k_1,p_a) \over 2k_1p_a} +eQ_e{b_{\sigma_1}(k_1,p_b) \over 2k_1p_b}\quad.
    \end{split}
\end{equation}
The soft part is now clearly separated from the remaining non-IR part,
used in the CEEX exponentiation for construction of ${\cal O} (\alpha)$
corrections. We have ordered the expression,
with the help of expansion similar to the contact interaction for $W$ 
propagator as well.
In $  \hat \Bmf\left[\st^{p}_{\lambda} \right]$ we
use an auxiliary fixed transfer $t_0$, independent of the place where
the photon is attached to the fermion line. In fact  $t_0$ is 
arbitrary and the choice $t_0=0$ could be used as well%
\footnote{The choice is nonetheless important from the point of view
of efficiency, it  affects size   
 of corrections in CEEX expansion scheme. Condition that
$\; \lim_{k_1 \to 0}  \; \; t_{a/b} =t_0$ is desirable.   
}. With the help of
$\bar{\Bmf}$ we provide the  residual  contribution calculated as 
a difference of the expression calculated with 
the true $t$-transfers ($t_a$ or $t_b$) and the auxiliary $t_0$ one. 
Note that ${\Bmf} = \hat{\Bmf} + \bar{\Bmf}$. Each of the
 contribution to
the sum given in the first equation of  (\ref{first-order-isr}) is independently gauge invariant.

We can see, that it was possible  to separate the complete spin amplitude for the process
$e^+ e^- \rightarrow \bar \nu_e \nu_e \gamma$
 into {\it six} individually QED gauge invariant parts. This conclusion is rather
straightforward to check, replacing photon polarization vector with its 
four-momentum. Each of the obtained parts  has rather well defined physical 
interpretation. 
It is also easy to verify that the  gauge invariance of each part
can be easily preserved to the case of the extrapolation, when because of 
additional photons, condition $p_a+p_b=p_c+p_d+k_1$ is not valid.
 Let us elaborate on this point a bit more.

\noindent $\bullet$
The first term of the soft photon type $\sfac^{\{I\}}_{\sigma_1}(k_1) 
  \hat  \Bmf\left[\st^{p}_{\lambda} \right]$ is gauge invariant thanks 
to invariance
of the standard ISR soft factor $\sfac^{\{I\}}_{\sigma_1}(k_1)  $. 
It is also of the universal form, identical for the diagrams with $s$-channel
$Z$-exchange as well as $t$-channel $W$.  

\noindent $\bullet$
The next two terms 
    $\bigl(  r_{\{I\}}^{B'} + {\cal M}^{1'} \bigr)$   
and $\bigl(  r_{\{I\}}^{B'} + {\cal M}^{1''} \bigr)$ 
originate only from diagrams of $t$-channel $W$ exchange. For the gauge 
invariance to hold, the  $t$-channel transfers 
have to be 
$t_a=(p_a -p_c)^2$, $t_b=(p_a -k_1-p_c)^2$  for the first term (and
$t_a=(p_b-k_1 -p_d)^2$, $t_b=(p_b -p_d)^2$ for the second one)\footnote{It is 
interesting to realize that only part of the
diagram is involved in the cancellation, namely  fermion and boson lines,
from which emission of the photon takes place. This observation 
will become useful in case of double bremsstrahlung amplitudes.}.

\noindent $\bullet$
The consecutive two terms $  r_{\{I\}}^{A'} \left( \st^{p}_{\lambda} \st^{k_1}_{\sigma_1} \right)$ 
and $ r_{\{I\}}^{A''} \left( \st^{p}_{\lambda} \st^{k_1}_{\sigma_1} \right)$ are again of the
same universal form as for any $s$-channel process and gauge invariant by construction.
These are also the terms which lead to leading-log (but not infrared) 
singular  terms, after phase space integration. In the 
corresponding Feynman diagrams the photon polarization vector and its 
momentum stand side
by side. 

\noindent $\bullet$
Finally, for the last expression 
$ \bigl({\cal M}^2+{\cal M}^3\bigr)$ to be gauge invariant it is enough that 
for the two terms choices for transfers $t_a$, $t_b$ are identical;  the same 
reduction procedure%
\footnote{Mechanism of gauge cancellation is fulfilled already 
at the level of this part of bosonic interaction alone. 
This observation will be useful in study of double bremsstrahlung amplitudes.}
is used. 

\vskip 2 mm
\centerline{ \bf Simplest case of $e^+e^- \to \nu_\mu \bar \nu_\mu$}
\vskip 2 mm

Let us finish this Section with the discussion of the
 $Z$ exchange part of the amplitude  (\ref{isr-feynman1}) in simple language of 
spinors and four-vectors. This part of the amplitude is important because it 
will define the framework for our main results collected  in Section 3.
\begin{equation}
  \label{isr-Z}
  \begin{split}
    \Meu^Z_{1\{I\}}\left( \st^{p}_{\lambda} \st^{k_1}_{\sigma_1} \right)=
   & eQ_e \;
    \bar{v}(p_b,\lambda_b)\; \mathbf{M}_{\{I\}}\;
         {\not\!{p_a}+m-{\not\!k_1} \over -2k_1p_a} \not\!{\epsilon}^\star_{\sigma_1}(k_1)\;
    u(p_a,\lambda_a)\\
   +&eQ_e \;
    \bar{v}(p_b,\lambda_b)
         \not\!{\epsilon}^\star_{\sigma_1}(k_1)\; {-{\not\!p_b}+m+\not\!{k_1} \over -2k_1p_b} \;
         \mathbf{M}_{\{I\}}\;
    u(p_a,\lambda_a) 
  \end{split}
\end{equation}
the superscript $ac$ or $bd$ can be dropped in $\mathbf{M}_{\{I\}}$ as it does not depend on the 
$t$ transfer of $W$ propagator, then $\mathbf{M}_{\{I\}}={\cal R}_Z$ of 
(\ref{cdotsZ}), see later in the text.  

The  gauge invariance of the two sub-parts proportional to 
 ${\not\!k_1}$ is straightforward to see, because the expression 
${\not\!k_1}\not\!{\epsilon}^\star_{\sigma_1}(k_1)$ alone is gauge invariant
thanks to ${\not\!k_1}{\not\!k_1}=0$. These parts of the amplitude do not contribute
to infrared singularity, however do contribute to the big logarithm related to
collinear singularity (once amplitudes are squared and integrated over the phase space). 
That is why, we will call these parts of the amplitude as  infrared finite collinear singular.
The remaining part of the amplitude:
\begin{equation}
  \label{isr-Zb}
  \begin{split}
    \Meu^{Z-ir}_{1\{I\}}\left( \st^{p}_{\lambda} \st^{k_1}_{\sigma_1} \right)=
   & eQ_e \;
    \bar{v}(p_b,\lambda_b)\; \mathbf{M}^{bd}_{\{I\}}\;
         {\not\!{p_a}+m \over -2k_1p_a} \not\!{\epsilon}^\star_{\sigma_1}(k_1)\;
    u(p_a,\lambda_a)\\
   +&eQ_e \;
    \bar{v}(p_b,\lambda_b)
         \not\!{\epsilon}^\star_{\sigma_1}(k_1)\; {-{\not\!p_b}+m \over -2k_1p_b} \;
         \mathbf{M}^{ac}_{\{I\}}\;
    u(p_a,\lambda_a) 
  \end{split}
\end{equation}

 factorizes  thanks to the orthogonality for Dirac spinors:
\begin{equation}
  \label{isr-Zc}
 \begin{split}
{\not\!{p_a}+m }   &= \sum_\lambda u(p_a,\lambda) \bar u(p_a,\lambda) \\
{- \not\!{p_b}+m } &= -\sum_\lambda v(p_b,\lambda) \bar v(p_b,\lambda).
 \end{split}
\end{equation}
into gauge invariant
soft photon factor 
 and Born amplitude:
\begin{equation}
  \begin{split}
    \Meu^{Z-ir}_{1\{I\}}\left( \st^{p}_{\lambda} \st^{k_1}_{\sigma_1} \right)
   &=\; \sfac^{\{I\}}_{\sigma_1}(k_1) \;
    \bar{v}(p_b,\lambda_b)\; \mathbf{M}^{bd}_{\{I\}}\;
    \;    u(p_a,\lambda_a);\\
   \sfac^{\{I\}}_{\sigma_1}(k_1)
&=\;+{eQ_e \over -2k_1p_a}\; 
\bar{u}(p_a,\lambda) \not\!{\epsilon}^\star_{\sigma_1}(k_1) \; u(p_a,\lambda_a) 
 +{eQ_e \over 2k_1p_b}\; 
\bar{v}(p_b,\lambda_b) \not\!{\epsilon}^\star_{\sigma_1}(k_1) \; v(p_b,\lambda) 
\\ 
&= 
 { -eQ_e \over 2k_1p_a }  b_{\sigma_1}(k_1,p_a)\delta_{\lambda\lambda_a}  
+{ eQ_e \over 2k_1p_b }  b_{\sigma_1}(k_1,p_b)\delta_{\lambda\lambda_b}  
  \end{split}
\end{equation}

The gauge invariance, takes place 
 in case of $Z$ exchange (and also $W$ exchange if approximation of contact
interaction is used) because then 
$\mathbf{M}_{\{I\}}=\mathbf{M}^{bd}_{\{I\}}=\mathbf{M}^{ac}_{\{I\}}$,  also two
parts of $\sfac^{\{I\}}_{\sigma_1}(k_1)$ are diagonal respectively
 in indices $\lambda\lambda_{a,b}$. 
The Born level spin amplitude factorizes out,
and the gauge dependent soft factors for emission from electron and positron lines, 
can be summed to gauge invariant
$\sfac^{\{I\}}_{\sigma_1}(k_1)$. For the explicit definition of $ b_{\sigma_1}(k_1,p_b)$,
see e.g. formula (231) of ref.\cite{Jadach:2000ir}.  
This part of the amplitude is infrared singular. 
We will use the factorization of the soft factors explained here, also later in the paper.

Note that in  case of single photon $Z$ exchange amplitude, we got only three
 gauge invariant parts: 
infrared-singular, and two others contributing to collinear-singular terms 
(after phase space integration). 
The residual (contributing  only non enhanced terms after phase space integration) terms are absent. 
Note however, that they are present in case of the $W$ exchange diagrams.

Finally let us comment that similar pattern of amplitude separation into gauge invariant 
parts can be observed for $ W^{\pm} \to l \nu_l \gamma$ in \cite{Nanava:2003cg}.

\section{Double bremsstrahlung}

In the present section we will study 
 the amplitudes for double bremsstrahlung in $e^+e^- \to \nu_e \bar \nu_e$ production
process. There are two classes of diagrams in this case, the first one with  $Z$ boson 
exchange in $s$ channel and the second one with $t$ channel $W$ exchange. Similarly
as in previous section and single bremsstrahlung  
we will look if  gauge invariant parts of the complete amplitude can be defined. 
We will be also interested,
if this can be done in a semi-automatic way,  
directly from the Feynman rules.  

The presentation of the complete amplitudes is rather difficult because 
of their length. 
To avoid lengthy formulas,
we will start with largely incomplete set of diagrams, nonetheless sufficient to localize
some gauge invariant group of terms. Once localized, 
it will be hidden under  symbol $L_a^b$ and, to the remaining part contributions  from the next
diagrams will be added. Again gauge invariant group of terms will be searched for.
 This procedure will be repeated until the complete list of diagrams 
of our process will be exhausted. The choice for the first diagram in this
procedure is motivated  by its particular (unique) form, later it is motivated
 by the form of the gauge dependent rest remaining from the previous step.  
For short hand notations we will use extended subscripts and superscripts for $L_a^b$.
For example we will  use symbol $L_{e^-}^{k_1,k_2}(n)$,
to denote contribution for the diagram with: first photon of momentum $k_1$ and second
$k_2$, attached to incoming $e^-$ line. The number $n$ in bracket (if present) will denote 
that it is only a part of 
the contribution from the particular Feynman diagram (or diagrams).
There will be often bar over this number to point that the particular part is
 gauge dependent.

Let us start our iteration  with diagrams involving double fermion propagator, that is 
diagrams where two photons are attached either to incoming electron or to incoming positron.
These are the {\it only} diagrams with  $k_1 \cdot k_2$ term in fermion propagators.
Our first aim will be thus to localize the parts which are gauge invariant by themselves, 
and include  this $k_1 \cdot k_2$ term. 
 Let us consider the eight diagrams with the photon lines attached
either to electron or positron line, see fig~\ref{doublee}. 
Explicitly, we will write down the part of the amplitude
corresponding to the incoming electron line only. In fact 
the diagrams with $Z$ and $W$ exchange are quite similar:

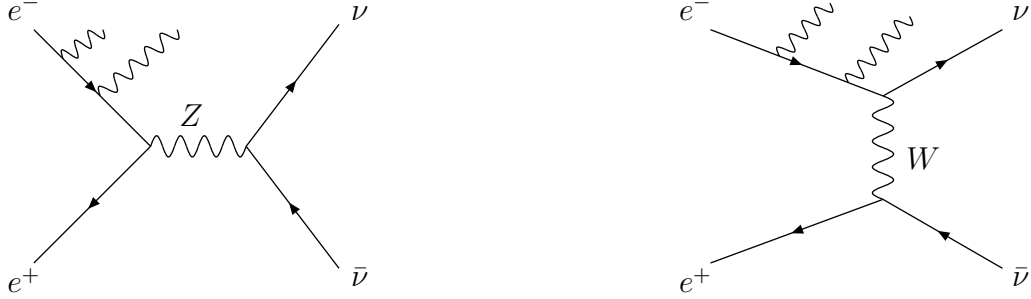
\begin{figure}[!ht] 
\begin{center}
 \begin{picture}(380,150)(-10,-30)
   \Photon(44,26)(80,26){4}{4}  \Text(55,42)[lt]{$Z$} 
   \ArrowLine(80,26)(115,72)   \Text(120,74)[lb]{${\nu}$}
   \ArrowLine(115,-20)(80,26)  \Text(120,-20)[lt]{$\bar{\nu}$}
   \ArrowLine(0,70)(44,26)   \Text(-10,74)[lb]{$e^-$}
   \ArrowLine(44,26)(0,-18)  \Text(-10,-20)[lt]{$e^+$}
   \Photon(11,59)(26,70){3}{3} 
   \Photon(25,45)(54,70){3}{5} 
   \Photon(320, 6)(320,45){4}{4}  \Text(330,26)[lt]{$W $}
   \ArrowLine(320,45)(365,70)   \Text(370,74)[lb]{$ {\nu} $}
   \ArrowLine(365,-20)(320, 6)  \Text(370,-20)[lt]{$ \bar{\nu}$}
   \ArrowLine(255,70)(320,45)   \Text(244,74)[lb]{$e^-$}
   \ArrowLine(320, 6)(255,-18)  \Text(244,-20)[lt]{$e^+$}
   \Photon(281,60)(300,80){3}{4} 
   \Photon(307,50)(330,75){3}{5} 

 \end{picture}
\end{center}
\caption{ Double emission from electron}
\label{doublee}  
\end{figure} 
\vspace{0.1cm}

\begin{equation}
  \label{isr-doublee}
  \begin{split}
    L_{e^-}^{k_1,k_2}= (eQ_e)^2  \bar v(p_b,\lambda_b) {\cal R}_B \Bigl( 
   &   
         {\not\!{p_a}+m-{\not\!k_1}-{\not\!k_2} \over -2k_1p_a -2k_2p_a -2k_1k_2} \not\!{\epsilon}^\star_{\sigma_2}(k_2)\;
         {\not\!{p_a}+m-{\not\!k_1} \over -2k_1p_a} \not\!{\epsilon}^\star_{\sigma_1}(k_1)\;
    u(p_a,\lambda_a)\\
   +& 
         {\not\!{p_a}+m-{\not\!k_1}-{\not\!k_2} \over -2k_1p_a -2k_2p_a -2k_1k_2} \not\!{\epsilon}^\star_{\sigma_1}(k_1)\;
         {\not\!{p_a}+m-{\not\!k_2} \over -2k_2p_a} \not\!{\epsilon}^\star_{\sigma_2}(k_2)\;
    u(p_a,\lambda_a) \Bigr)
  \end{split}
\end{equation}

 Expression ${\cal R}_B$ describing final state neutrino interaction and $Z$ or $W$ exchange, will be defined
later, respectively in formulas (\ref{cdotsZ}) and (\ref{cdotsW}). We can separate formula (\ref{isr-doublee}) into the following parts 
\begin{equation}
\label{isr-doublees}
 L_{e^-}^{k_1,k_2}= L_{e^-}^{k_1,k_2}(1)+ L_{e^-}^{k_1,k_2}(2)+L_{e^-}^{k_1,k_2}(\bar 3)+ L_{e^-}^{k_1,k_2}(\bar 4),
\end{equation} 
where:
\begin{equation}
  \label{isr-doubleea}
  \begin{split}
    L_{e^-}^{k_1,k_2}(1)= (eQ_e)^2  \bar v(p_b,\lambda_b)  {\cal R}_B\Bigl( 
   &   
         {-{\not\!k_2} \over -2k_1p_a -2k_2p_a -2k_1k_2} \not\!{\epsilon}^\star_{\sigma_2}(k_2)\;
         {-{\not\!k_1} \over -2k_1p_a} \not\!{\epsilon}^\star_{\sigma_1}(k_1)\;
    u(p_a,\lambda_a)\\
   +& 
         {-{\not\!k_1} \over -2k_1p_a -2k_2p_a -2k_1k_2} \not\!{\epsilon}^\star_{\sigma_1}(k_1)\;
         {-{\not\!k_2} \over -2k_2p_a} \not\!{\epsilon}^\star_{\sigma_2}(k_2)\;
    u(p_a,\lambda_a) \Bigr)
  \end{split}
\end{equation}
is gauge invariant by construction,  thanks
to terms   ${\not\!k_1}\not\!{\epsilon}^\star_{\sigma_1}(k_1)$
and  ${\not\!k_2}\not\!{\epsilon}^\star_{\sigma_2}(k_2)$ which 
are gauge invariant by themselves. This is similar to the case of single bremsstrahlung.
 The second part
\begin{equation}
  \label{isr-doubleeb}
  \begin{split}
    L_{e^-}^{k_1,k_2}(2)= (eQ_e)^2  \bar v(p_b,\lambda_b) {\cal R}_B \Bigl( 
   &   
         {\not\!{p_a}+m-{\not\!k_2} \over -2k_1p_a -2k_2p_a -2k_1k_2} \not\!{\epsilon}^\star_{\sigma_1}(k_1)\;
         {-{\not\!k_2} \over -2k_2p_a} \not\!{\epsilon}^\star_{\sigma_2}(k_2)\;
    u(p_a,\lambda_a)\\
   +& 
         {-{\not\!k_2} \over -2k_1p_a -2k_2p_a -2k_1k_2} \not\!{\epsilon}^\star_{\sigma_1}(k_1)\;
         {\not\!{p_a}+m \over -2k_2p_a} \not\!{\epsilon}^\star_{\sigma_2}(k_2)\;
    u(p_a,\lambda_a) 
\\
   -& 
         {\not\!k_2}\Bigl({1 \over -2k_1p_a -2k_2p_a -2k_1k_2} -{1 \over -2k_2p_a  }\Bigr)\\
   & \not\!{\epsilon}^\star_{\sigma_2}(k_2)\;
         {\not\!{p_a}+m \over -2k_1p_a} \not\!{\epsilon}^\star_{\sigma_1}(k_1)\;
    u(p_a,\lambda_a) \\
   &   
        + (\not\!{p_a}+m )\Bigl({1 \over -2k_1p_a -2k_2p_a -2k_1k_2} -{1 \over -2k_1p_a -2k_2p_a }\Bigr) \\
   &       \not\!{\epsilon}^\star_{\sigma_2}(k_2)\;
         {\not\!{p_a}+m \over -2k_1p_a} \not\!{\epsilon}^\star_{\sigma_1}(k_1)\;
    u(p_a,\lambda_a)
     \Bigr)\\
   &  + (1 \leftrightarrow 2)
  \end{split}
\end{equation}
is also gauge invariant, but it need to be checked by direct calculation. 
This contribution, like the previous one, is free of infrared singularity. 
In definition of $ L_{e^-}^{k_1,k_2}(2)$ we had to introduce subtraction; 
terms proportional to ${\not\!k_2}\bigl( -{1 \over -2k_2p_a  }\bigr)$ and 
$ (\not\!{p_a}+m ) \bigl(  -{1 \over -2k_1p_a -2k_2p_a } \bigr) $
The subtraction terms are added back to 
(\ref{isr-doublees}) as formulas  (\ref{isr-doubleec}) and (\ref{isr-doubleed}),
 but with the opposite sign of course. 
It is important to realize,
that these  subtraction terms are defined uniquely by the $Z$ exchange part of the  
amplitude for single photon emission\footnote{The subtraction term 
for the $W$ exchange differs only by the replacement ${\cal R}_B= {\cal R}_W$.},
see formula (\ref{isr-Z}).
It has  to be multiplied by the 
soft photon factor for the second photon and separated into ifrared finite and infrared
singular parts, as explained in section 2.

\begin{equation}
  \label{isr-doubleec}
  \begin{split}
    L_{e^-}^{k_1,k_2}(\bar 3)= (eQ_e)^2  \bar v(p_b,\lambda_b) {\cal R}_B \Bigl( 
   &   
         {\not\!{p_a}+m \over -2k_1p_a -2k_2p_a } \not\!{\epsilon}^\star_{\sigma_2}(k_2)\;
         {\not\!{p_a}+m \over -2k_1p_a} \not\!{\epsilon}^\star_{\sigma_1}(k_1)\;
    u(p_a,\lambda_a)\\
   & 
        + {\not\!{p_a}+m \over -2k_1p_a -2k_2p_a } \not\!{\epsilon}^\star_{\sigma_1}(k_1)\;
         {\not\!{p_a}+m \over -2k_2p_a} \not\!{\epsilon}^\star_{\sigma_2}(k_2)\;
    u(p_a,\lambda_a)  \Bigr)\\
= (eQ_e)^2  \bar v(p_b,\lambda_b) {\cal R}_B
   &   
         {\not\!{p_a}+m \over -2k_2p_a} \not\!{\epsilon}^\star_{\sigma_2}(k_2)\;
         {\not\!{p_a}+m \over -2k_1p_a} \not\!{\epsilon}^\star_{\sigma_1}(k_1)\;
    u(p_a,\lambda_a)
  \end{split}
\end{equation}

 The next term $ L_{e^-}^{k_1,k_2}(\bar 4)$ is, also free of $k_1k_2$
and its numerator is linear in the photon momentum
\begin{equation}
  \label{isr-doubleed}
  \begin{split}
    L_{e^-}^{k_1,k_2}(\bar 4)= (eQ_e)^2  \bar v(p_b,\lambda_b) {\cal R}_B \Bigl( 
   &   
         {-{\not\!k_2} \over -2k_2p_a  } \not\!{\epsilon}^\star_{\sigma_2}(k_2)\;
         {\not\!{p_a}+m \over -2k_1p_a} \not\!{\epsilon}^\star_{\sigma_1}(k_1)\;
    u(p_a,\lambda_a)\\
   +& 
         {-{\not\!k_1} \over -2k_1p_a } \not\!{\epsilon}^\star_{\sigma_1}(k_1)\;
         {\not\!{p_a}+m \over -2k_2p_a} \not\!{\epsilon}^\star_{\sigma_2}(k_2)\;
    u(p_a,\lambda_a) \Bigr).
  \end{split}
\end{equation}

The complete contribution from diagrams of fig.~\ref{doublee},
formula(\ref{isr-doublees}),  is not gauge invariant. The last two terms 
are  gauge dependent, but are also relatively short. The first one (\ref{isr-doubleec}) 
has a structure of Born amplitude multiplied by soft photon factors, 
 the second one (\ref{isr-doubleed}), has a structure of soft photon emission
for one of the two photons only, see discussion at the end of Section 2.

\begin{figure}[!ht] 
\begin{center}
 \begin{picture}(380,150)(-10,-30)
   \Photon(44,26)(80,26){4}{4}  \Text(55,42)[lt]{$Z$} 
   \ArrowLine(80,26)(115,72)   \Text(120,74)[lb]{${\nu}$}
   \ArrowLine(115,-20)(80,26)  \Text(120,-20)[lt]{$\bar{\nu}$}
   \ArrowLine(0,70)(44,26)   \Text(-10,74)[lb]{$e^-$}
   \ArrowLine(44,26)(0,-18)  \Text(-10,-20)[lt]{$e^+$}
   \Photon(11,59)(26,70){3}{3} 
   \Photon(11,-7)(26,-18){3}{3} 
   \Photon(320, 6)(320,45){4}{4}  \Text(330,26)[lt]{$W $}
   \ArrowLine(320,45)(365,70)   \Text(370,74)[lb]{$ {\nu} $}
   \ArrowLine(365,-20)(320, 6)  \Text(370,-20)[lt]{$ \bar{\nu}$}
   \ArrowLine(255,70)(320,45)   \Text(244,74)[lb]{$e^-$}
   \ArrowLine(320, 6)(255,-18)  \Text(244,-20)[lt]{$e^+$}
   \Photon(281,60)(300,80){3}{4} 
   \Photon(281,-9)(300,-18){3}{4} 

 \end{picture}
\end{center}
\caption{ Single emission from electron and positron}
\label{sinsin}  
\end{figure}
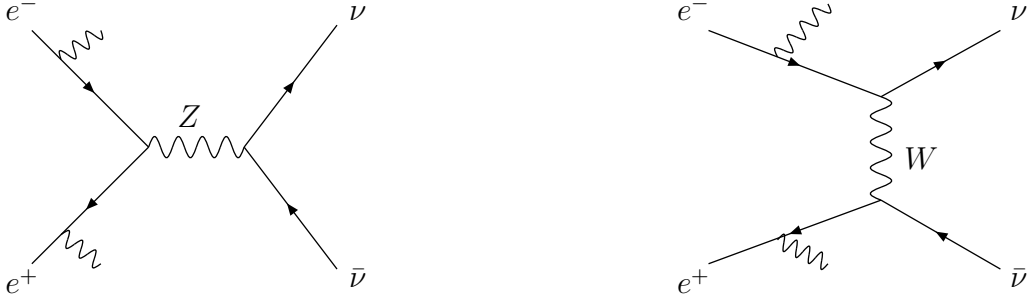

Once we have completed the diagrams with two photon lines attached to one fermion line, let us turn to another group of diagrams,
where one of the photons is attached to electron and another one to positron line, see Fig.~\ref{sinsin}.
 Note that for subgroup of diagrams with  
$Z$ boson exchange  only, these are the last contributing diagrams:
\begin{equation}
  \label{isr-sinsine}
  \begin{split}
    L_{e^-,e^+}^{k_1,k_2}= (eQ_e)^2 \Bigl( 
    &   \bar v(p_b,\lambda_b)   \not\!{\epsilon}^\star_{\sigma_2}(k_2)      {-\not\!{p_b}+m+{\not\!k_2} \over -2k_2p_b } \;  {\cal R}_B
         {\not\!{p_a}+m-{\not\!k_1} \over -2k_1p_a} \not\!{\epsilon}^\star_{\sigma_1}(k_1)\;
    u(p_a,\lambda_a)\\
   +&    \bar v(p_b,\lambda_b) \not\!{\epsilon}^\star_{\sigma_1}(k_1) {-\not\!{p_b}+m+{\not\!k_1} \over -2k_1p_b} \;  {\cal R}_B
         {\not\!{p_a}+m-{\not\!k_2} \over -2k_2p_a} \not\!{\epsilon}^\star_{\sigma_2}(k_2)\;
    u(p_a,\lambda_a) \Bigr)
  \end{split}
\end{equation}

\vspace{0.1cm}

As in the previous case the expression for $L_{e^-,e^+}^{k_1,k_2}$ can be easily separated into parts,
\begin{equation}
  \label{isr-sinsinex}
L_{e^-,e^+}^{k_1,k_2}=L_{e^-,e^+}^{k_1,k_2}(1)+L_{e^-,e^+}^{k_1,k_2}(\bar 2)+L_{e^-,e^+}^{k_1,k_2}(\bar 3)
\end{equation}
The first part $L_{e^-,e^+}^{k_1,k_2}(1)$ 
is by construction gauge invariant, it is also the only part from this group of diagrams with  numerator  proportional both to the
momenta of $k_1$ and $k_2$:
\begin{equation}
  \label{isr-sinsinea}
  \begin{split}
    L_{e^-,e^+}^{k_1,k_2}(1)= (eQ_e)^2 \Bigl( 
    &   \bar v(p_b,\lambda_b)   \not\!{\epsilon}^\star_{\sigma_2}(k_2)      {{\not\!k_2} \over -2k_2p_b } \; {\cal R}_B
         {-{\not\!k_1} \over -2k_1p_a} \not\!{\epsilon}^\star_{\sigma_1}(k_1)\;
    u(p_a,\lambda_a)\\
   +&    \bar v(p_b,\lambda_b) \not\!{\epsilon}^\star_{\sigma_1}(k_1) {{\not\!k_1} \over -2k_1p_b} \;  {\cal R}_B
         {-{\not\!k_2} \over -2k_2p_a} \not\!{\epsilon}^\star_{\sigma_2}(k_2)\;
    u(p_a,\lambda_a) \Bigr)
  \end{split}
\end{equation}
The second term $L_{e^-,e^+}^{k_1,k_2}(\bar 2)$ has two contributions, of numerators linear either in  $k_1$ or  $k_2$, it reads:
\begin{equation}
  \label{isr-sinsineb}
  \begin{split}
    L_{e^-,e^+}^{k_1,k_2}(\bar 2)= (eQ_e)^2 \Bigl( 
    &   \bar v(p_b,\lambda_b)   \not\!{\epsilon}^\star_{\sigma_2}(k_2)      {-\not\!{p_b}+m \over -2k_2p_b } \;  {\cal R}_B
         {-{\not\!k_1} \over -2k_1p_a} \not\!{\epsilon}^\star_{\sigma_1}(k_1)\;
    u(p_a,\lambda_a)\\
   +&    \bar v(p_b,\lambda_b) \not\!{\epsilon}^\star_{\sigma_1}(k_1) {{\not\!k_1} \over -2k_1p_b} \;  {\cal R}_B
         {\not\!{p_a}+m \over -2k_2p_a} \not\!{\epsilon}^\star_{\sigma_2}(k_2)\;
    u(p_a,\lambda_a) \\
    +&   \bar v(p_b,\lambda_b)   \not\!{\epsilon}^\star_{\sigma_2}(k_2)      {{\not\!k_2} \over -2k_2p_b } \;  {\cal R}_B
         {\not\!{p_a}+m \over -2k_1p_a} \not\!{\epsilon}^\star_{\sigma_1}(k_1)\;
    u(p_a,\lambda_a)\\
   +&    \bar v(p_b,\lambda_b) \not\!{\epsilon}^\star_{\sigma_1}(k_1) {-\not\!{p_b}+m \over -2k_1p_b} \; {\cal R}_B 
         {-{\not\!k_2} \over -2k_2p_a} \not\!{\epsilon}^\star_{\sigma_2}(k_2)\;
    u(p_a,\lambda_a) \Bigr)
  \end{split}
\end{equation}
Finally the third one  $L_{e^-,e^+}^{k_1,k_2}(\bar 3)$ is free of the $k_1$ or  $k_2$ in the numerator.
\begin{equation}
  \label{isr-sinsinec}
  \begin{split}
    L_{e^-,e^+}^{k_1,k_2}(\bar 3)= (eQ_e)^2 \Bigl( 
    &   \bar v(p_b,\lambda_b)   \not\!{\epsilon}^\star_{\sigma_2}(k_2)      {-\not\!{p_b}+m \over -2k_2p_b } \;  {\cal R}_B
         {\not\!{p_a}+m \over -2k_1p_a} \not\!{\epsilon}^\star_{\sigma_1}(k_1)\;
    u(p_a,\lambda_a)\\
   +&    \bar v(p_b,\lambda_b) \not\!{\epsilon}^\star_{\sigma_1}(k_1) {-\not\!{p_b}+m \over -2k_1p_b} \;  {\cal R}_B
         {\not\!{p_a}+m \over -2k_2p_a} \not\!{\epsilon}^\star_{\sigma_2}(k_2)\;
    u(p_a,\lambda_a) \Bigr)
  \end{split}
\end{equation}

To complete the sub-set of diagrams for double bremsstrahlung from 
initial state, the contribution of the double emission
from positron line should be added. We will omit explicit formulas here,
the explicit expressions for $L_{e^+}^{k_1,k_2}(1)$, $ L_{e^+}^{k_1,k_2}(2)$, $L_{e^+}^{k_1,k_2}(\bar 3)$, $ L_{e^+}^{k_1,k_2}(\bar 4)$ can be obtained from
 $ L_{e^-}^{k_1,k_2}(1)$, $ L_{e^-}^{k_1,k_2}(2$, $)L_{e^-}^{k_1,k_2}(\bar 3)$, $ L_{e^-}^{k_1,k_2}(\bar 4) $ by analogy or explicit calculation. 

\subsection{Diagrams with $Z$ exchange}

Before going to the more complex case of $W$ exchange, where complications due to $t$ dependence
of $W$ propagator occur, let us concentrate on the simpler one,  $Z$ exchange alone. The diagrams
discussed so far, represent then the complete gauge invariant amplitude
for the process  $e^+e^- \to \nu_\mu \bar \nu_\mu \gamma \gamma$.
In such a sub-group of diagrams (for the process $e^+e^- \to \nu_e \bar \nu_e \gamma \gamma$)
symbol ${\cal R}_{B=Z} $ represents always

\begin{equation}
  \label{cdotsZ}
{\cal R}_Z =   \Bigl(\gamma^\mu(v {\bf 1} + a \gamma^5)\Bigr)_{\alpha\beta}  \; \; \Bigl( \bar u(p_c,\lambda_c) \gamma_\mu(v {\bf 1} + a \gamma^5)  v(p_d,\lambda_d)\Bigr)BW_Z((p_c+p_d)^2)
\end{equation}
which is a constant algebraic expression, independent on photon momenta and identical for all diagrams listed. 
The $Z$ boson propagator $BW_Z((p_c+p_d)^2)$ depends on the invariant mass of the outcoming neutrinos only.
The bi-spinor indices of $ \gamma^\mu$, $ \gamma^\mu\gamma^5$ matrices which enter into the matrix products
of  formulae such as  (\ref{isr-doublee})   to (\ref{isr-sinsinec}) are explicitly given as $\alpha\beta$.
The complete amplitude reads:
\begin{equation}
 \label{Zgg-a}
  \begin{split}
\cal{M}=& L_{e^-}^{k_1,k_2}+ L_{e^+}^{k_1,k_2}+ L_{e^-,e^+}^{k_1,k_2} \\
       =&L_{e^-}^{k_1,k_2}(1)+ L_{e^-}^{k_1,k_2}(2)+L_{e^-}^{k_1,k_2}(\bar 3)+ L_{e^-}^{k_1,k_2}(\bar 4)+\\
        &L_{e^+}^{k_1,k_2}(1)+ L_{e^+}^{k_1,k_2}(2)+L_{e^+}^{k_1,k_2}(\bar 3)+ L_{e^+}^{k_1,k_2}(\bar 4)+\\
        &L_{e^-,e^+}^{k_1,k_2}(1)+L_{e^-,e^+}^{k_1,k_2}(\bar 2)+L_{e^-,e^+}^{k_1,k_2}(\bar 3).
  \end{split}
\end{equation}

Where $ L_{e^-}^{k_1,k_2}$  
is given by formula (\ref{isr-doublee}) (or if separated into parts by (\ref{isr-doublees})) and $L_{e^-,e^+}^{k_1,k_2}$
by (\ref{isr-sinsine}) (or (\ref{isr-sinsinex})).
For $L_{e^+}^{k_1,k_2}$ the expressions of 
 $ L_{e^-}^{k_1,k_2}$ can be used, with appropriate changes of signs momenta etc.

The formula for the complete spin amplitude ($Z$ exchange only) can be easily re-ordered
into consecutive contributions ${\cal M}_1, {\cal M}_2, {\cal M}_3, ...$, each gauge invariant, and each expressed 
as  group of $L$'s in the bracket or just individual  $L$:

\begin{equation}
 \label{Zgg-b}
  \begin{split}
 {\cal M}=& \Meu^Z_{2\{I\}}\left( \st^{p}_{\lambda} \st^{k_1}_{\sigma_1}\st^{k_2}_{\sigma_2} \right)\\
=& {\cal M}_1+ {\cal M}_2+ {\cal M}_3+  {\cal M}_4+ {\cal M}_5+ {\cal M}_6+ {\cal M}_7\\
       =&L_{e^-}^{k_1,k_2}(1)+ L_{e^-}^{k_1,k_2}(2)+ 
         L_{e^+}^{k_1,k_2}(1)+ L_{e^+}^{k_1,k_2}(2)+ 
         L_{e^-,e^+}^{k_1,k_2}(1) \\
         +&\bigr(L_{e^-}^{k_1,k_2}(\bar 4)+L_{e^+}^{k_1,k_2}(\bar 4)+L_{e^-,e^+}^{k_1,k_2}(\bar 2) \bigl)
         +\bigr((L_{e^-}^{k_1,k_2}(\bar 3)+L_{e^+}^{k_1,k_2}(\bar 3)+L_{e^-,e^+}^{k_1,k_2}(\bar 3) \bigl)
  \end{split}
\end{equation}

As one can see, the sum of terms ${\cal M}_1$  to ${\cal M}_5$, contributing to $\beta^2_2$ of the CEEX exponentiation scheme
(these terms are not infrared singular at all), is gauge invariant and 
clearly separated from the rest.  It can be even further sliced into
{\it five} parts, each individually gauge invariant. The last two terms, ${\cal M}_6$  and ${\cal M}_7$ 
corresponds respectively to $\beta^1_1$ and   $\beta^0_0$ (multiplied by one or two soft photon factors) 
and can be obtained from lower order of perturbation expansion. 
It is rather straigtforward to see, that the term
${\cal M}_7$ consist of Born level amplitude multiplied by soft factors corresponding to emission 
of two photons.
The term  ${\cal M}_6$ can be seen as consisting of two factors; for one of  the photons  soft factor, and for the other one
term of $\beta^1_1$. To see it better it is convenient  to order the expression accordingly to terms
proportional either to  ${\not\!k_1}$ or  ${\not\!k_2}$.

Note also, that for the each of the parts to be gauge invariant,  
it is not necessary that four-momentum conservation is fulfiled.
That is  why, the separation is easily adaptable to extrapolation procedure as used in KKMC.

This completes our discussion of results for $s$-channel exchange of $Z$.
Let us now turn to the contributions related to the $t$-channel $W$-exchange.

\subsection{Diagrams with $W$ exchange}
First, let us note that all formulae presented so far 
are valid for the diagrams involving $W$ exchange as well. The difference is that instead of formula (\ref{cdotsZ}) for 
${\cal R}_{B}$  one should use:
\begin{equation}
  \label{cdotsW}
{\cal R}_W =  \Bigl(  \gamma_\mu( {\bf 1} -\gamma^5)  v(p_d,\lambda_d)\Bigr)_{\alpha}  \; \;  \Bigl(\bar u(p_c,\lambda_c)\gamma^\mu( {\bf 1} -\gamma^5)\Bigr)_{\beta}  BW_W(t)
\end{equation}
The spinorial form of this expression is universal, and  as in the case of $Z$ exchange, in all places the same 
expression is to be used. The difference
lies in $t$ dependence of $W$ propagator, the transfer 
will  depend on the way how the photon lines are attached to the fermion ones.
Nonetheless in some groups of terms gauge cancellation do occur anyway in the same way as before. 
If we recall the part of the $W$ exchange amplitude, written in analogy to (\ref{Zgg-b}) as:
\begin{equation}
 \label{Wgg-a}
  \begin{split}
\cal{M}_W^A=& {\cal M}_1+ {\cal M}_2+ {\cal M}_3+  {\cal M}_4+ {\cal M}_5+ \bar{\cal M}_6+ \bar{\cal M}_7\\
       =&L_{e^-}^{k_1,k_2}(1)+ L_{e^-}^{k_1,k_2}(2)+ 
         L_{e^+}^{k_1,k_2}(1)+ L_{e^+}^{k_1,k_2}(2)+ 
         L_{e^-,e^+}^{k_1,k_2}(1) \\
        & +\bigr(L_{e^-}^{k_1,k_2}(\bar 4)+L_{e^+}^{k_1,k_2}(\bar 4)+L_{e^-,e^+}^{k_1,k_2}(\bar 2) \bigl)
         +\bigr((L_{e^-}^{k_1,k_2}(\bar 3)+L_{e^+}^{k_1,k_2}(\bar 3)+L_{e^-,e^+}^{k_1,k_2}(\bar 3) \bigl)
  \end{split}
\end{equation}

then the parts ${\cal M}_1$, $ {\cal M}_2 $, ${\cal M}_3 $, $ {\cal M}_4 $ and  ${\cal M}_5$ remain
gauge invariant. Only the last two 
terms  will need contributions from diagrams with 
triple and quatric gauge boson
couplings for the gauge invariance to hold. To visualize this point the bar sign is now placed
over $\bar{\cal M}_6 $ and  $\bar{\cal M}_7$ .
Note, that as already pointed in the previous subsection
these are the contributions, which could be obtained (extra soft photon factors are only needed) 
from the results of the calculation at lower perturbative order if the 
complications due to variation of $W$ exchange transfers were not taken into account.

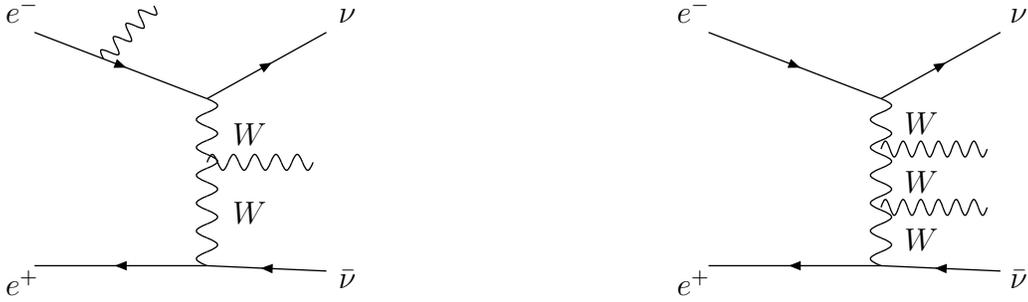
\begin{figure}[!ht] 
\begin{center}
 \begin{picture}(380,150)(-10,-30)
   \Photon(66, -18)(66,45){4}{6}  \Text(76,36)[lt]{$W $}  \Text(76, 6)[lt]{$W $}
   \ArrowLine(66,45)(111,70)   \Text(116,74)[lb]{$ {\nu} $}
   \ArrowLine(111,-20)(66, -18)  \Text(116,-20)[lt]{$ \bar{\nu}$}
   \ArrowLine(1,70)(66,45)   \Text(-10,74)[lb]{$e^-$}
   \ArrowLine(66, -18)(1,-18)  \Text(-10,-20)[lt]{$e^+$}
   \Photon(27,60)(46,80){3}{4} 
   \Photon(66,21)(106,21){3}{5} 
   \Photon(320, -18)(320,45){4}{6}  \Text(330,40)[lt]{$W $}  \Text(330, 18)[lt]{$W $} \Text(330, -4)[lt]{$W $}
   \ArrowLine(320,45)(365,70)   \Text(370,74)[lb]{$ {\nu} $}
   \ArrowLine(365,-20)(320, -18)  \Text(370,-20)[lt]{$ \bar{\nu}$}
   \ArrowLine(255,70)(320,45)   \Text(244,74)[lb]{$e^-$}
   \ArrowLine(320, -18)(255,-18)  \Text(244,-20)[lt]{$e^+$}
   \Photon(320,26)(360,26){3}{6} 
   \Photon(320, 4)(360, 4){3}{6} 
%
 \end{picture}
\end{center}
\caption{ Single and double emission from W}
\label{Wemi}  
\end{figure} 
\vspace{0.1cm}

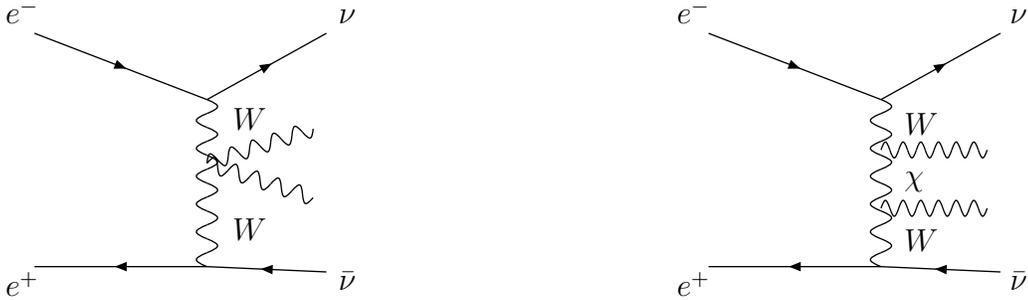
\begin{figure}[!ht] 
\begin{center}
 \begin{picture}(380,150)(-10,-30)
   \Photon(66, -18)(66,45){4}{6}  \Text(76,42)[lt]{$W $}  \Text(76, 0)[lt]{$W $}
   \ArrowLine(66,45)(111,70)   \Text(116,74)[lb]{$ {\nu} $}
   \ArrowLine(111,-20)(66, -18)  \Text(116,-20)[lt]{$ \bar{\nu}$}
   \ArrowLine(1,70)(66,45)   \Text(-10,74)[lb]{$e^-$}
   \ArrowLine(66, -18)(1,-18)  \Text(-10,-20)[lt]{$e^+$}
   \Photon(66,21)(106,34){3}{5} 
   \Photon(66,21)(106, 8){3}{5} 
   \Photon(320, -18)(320,45){4}{6}  \Text(330,40)[lt]{$W $}  \Text(330, 18)[lt]{$\chi $} \Text(330, -4)[lt]{$W $}
   \ArrowLine(320,45)(365,70)   \Text(370,74)[lb]{$ {\nu} $}
   \ArrowLine(365,-20)(320, -18)  \Text(370,-20)[lt]{$ \bar{\nu}$}
   \ArrowLine(255,70)(320,45)   \Text(244,74)[lb]{$e^-$}
   \ArrowLine(320, -18)(255,-18)  \Text(244,-20)[lt]{$e^+$}
   \Photon(320,26)(360,26){3}{6} 
   \Photon(320, 4)(360, 4){3}{6} 
%
 \end{picture}
\end{center}
\caption{ Four boson coupling and coupling for unphysical $\chi$ field.}
\label{Hemi}  
\end{figure} 
\vspace{0.1cm}

Let us continue with the second part of our discusion now.  
The two (now gauge dependent),
contributions $\bar{\cal M}_6 $ and  $\bar{\cal M}_7$ will be first 
 completed with 
 diagrams from the left-hand side of fig. \ref{Wemi}. 
Note that these diagrams are the  last ones with photon line attached 
to incoming electron/positron,
thus the last ones contributing  collinear and/or soft singularities. 
The contribution to the scattering amplitude from  these new diagrams 
reads:
\begin{equation}
  \label{isr-W}
  \begin{split}
    L_{e^-,W}^{k_1,k_2}= & (eQ_e)^2 BW_W\bigl((p_c+k_2-p_a)^2\bigr)  BW_W\bigl((p_c+k_2+k_1-p_a)^2\bigr)\\
\Bigl( &
      \bar v(p_b,\lambda_b)   \; ({\bf 1} - \gamma^5) \gamma^\mu  \;
        \;
    v(p_d,\lambda_d)\\ 
 &   \bigl[  
    g_{\mu\nu}(p-q)_\rho +g_{\nu\rho}(q-k_1)_\mu +g_{\mu\rho}(k_1-p)_\nu 
    \bigr] \bigl( {\epsilon}^\star_{\sigma_1}(k_1) \bigr)^\rho \\
    &    \bar u(p_c,\lambda_c)  \;  ({\bf 1} - \gamma^5) \gamma^\nu \;
         {\not\!{p_a}+m -{\not\!k_2} \over -2k_2p_a} \not\!{\epsilon}^\star_{\sigma_2}(k_2)\;
    u(p_a,\lambda_a)  \Bigr)\\
   &  + (1 \leftrightarrow 2)
  \end{split}
\end{equation}
Here  $p=p_d-p_b=-(p_c-p_a+k_1+k_2)$ and $q=p_c-p_a+k_2=-(p_d-p_b+k_1)$.  Similarily one can write contribution
$ L_{e^+,W}^{k_1,k_2}$ for the two diagrams with emission from positron and $W$, we will omit the corresponding
formulae.
As before, we separate $ L_{e^-,W}^{k_1,k_2}= L_{e^-,W}^{k_1,k_2}(k^0)+ L_{e^-,W}^{k_1,k_2}(k^1)$ into parts, $(k^1)$ marks contribution where  only 
${\not\!k_2}$ is taken from the fermionic propagator 
and  $(k^0)$ marks the rest.  The explicit 
formulas are:

\begin{equation}
  \label{isr-W-0}
  \begin{split}
    L_{e^-,W}^{k_1,k_2}(k^0)= & (eQ_e)^2 BW_W\bigl((p_c+k_2-p_a)^2\bigr)  BW_W\bigl((p_c+k_2+k_1-p_a)^2\bigr)\\ 
\Bigl( 
    &   \bar v(p_b,\lambda_b)   \; ({\bf 1} - \gamma^5) \gamma^\mu  \;
        \;
    v(p_d,\lambda_d)\\ 
 &   \bigl[  
    g_{\mu\nu}(p-q)_\rho +g_{\nu\rho}(q-k_1)_\mu +g_{\mu\rho}(k_1-p)_\nu 
    \bigr] \bigl( {\epsilon}^\star_{\sigma_1}(k_1) \bigr)^\rho \\
    &    \bar u(p_c,\lambda_c)  \;  ({\bf 1} - \gamma^5) \gamma^\nu \;
         {\not\!{p_a}+m  \over -2k_2p_a} \not\!{\epsilon}^\star_{\sigma_2}(k_2)\;
    u(p_a,\lambda_a)  \Bigr)\\
   &  + (1 \leftrightarrow 2)
  \end{split}
\end{equation}

and 
\begin{equation}
  \label{isr-W-k}
  \begin{split}
    L_{e^-,W}^{k_1,k_2}(k^1)= & (eQ_e)^2 BW_W\bigl((p_c+k_2-p_a)^2\bigr)  BW_W\bigl((p_c+k_2+k_1-p_a)^2\bigr)\\
\Bigl( 
    &   \bar v(p_b,\lambda_b)   \; ({\bf 1} - \gamma^5) \gamma^\mu  \;
        \;
    v(p_d,\lambda_d)\\ 
 &   \bigl[  
    g_{\mu\nu}(p-q)_\rho +g_{\nu\rho}(q-k_1)_\mu +g_{\mu\rho}(k_1-p)_\nu 
    \bigr] \bigl( {\epsilon}^\star_{\sigma_1}(k_1) \bigr)^\rho \\
    &    \bar u(p_c,\lambda_c)  \;  ({\bf 1} - \gamma^5) \gamma^\nu \;
         {-{\not\!k_2} \over -2k_2p_a} \not\!{\epsilon}^\star_{\sigma_2}(k_2)\;
    u(p_a,\lambda_a)  \Bigr)\\
   &  + (1 \leftrightarrow 2)
  \end{split}
\end{equation}

Let us start with the second one, which can be easily transformed (with help of Dirac equation) into 

\begin{equation}
  \label{isr-W-ka}
  \begin{split}
    L_{e^-,W}^{k_1,k_2}(k^1)= & (eQ_e)^2 BW_W\bigl((p_c+k_2-p_a)^2\bigr)  BW_W\bigl((p_c+k_2+k_1-p_a)^2\bigr)\\
\Bigl( 
    &   \bar v(p_b,\lambda_b)   \; ({\bf 1} - \gamma^5) \gamma^\mu  \;
        \;
    v(p_d,\lambda_d)\\ 
 &   \bigl[  
    g_{\mu\nu}(p_d-p_b-p_c+p_a-k_2)_\rho +g_{\nu\rho}(-2k_1)_\mu +g_{\mu\rho}(2k_1-p_a)_\nu 
    \bigr] \bigl( {\epsilon}^\star_{\sigma_1}(k_1) \bigr)^\rho \\
    &    \bar u(p_c,\lambda_c)  \;  ({\bf 1} - \gamma^5) \gamma^\nu \;
         {-{\not\!k_2} \over -2k_2p_a} \not\!{\epsilon}^\star_{\sigma_2}(k_2)\;
    u(p_a,\lambda_a)  \Bigr)\\
   &  + (1 \leftrightarrow 2).
  \end{split}
\end{equation}

This contribution can be separated even further:
\begin{equation}
  \label{isr-W-kk}
  \begin{split}
    L_{e^-,W}^{k_1,k_2}(k^1)=L_{e^-,W}^{k_1,k_2}(\bar 1)+L_{e^-,W}^{k_1,k_2}(2)+L_{e^-,W}^{k_1,k_2}(\bar 3)
  \end{split}
\end{equation}
where
\begin{equation}
  \label{isr-W-kb}
  \begin{split}
    L_{e^-,W}^{k_1,k_2}(\bar 1)= & (eQ_e)^2 BW_W\bigl((p_c+k_2-p_a)^2\bigr)  BW_W\bigl((p_c+k_2+k_1-p_a)^2\bigr)\\
\Bigl( 
    &   \bar v(p_b,\lambda_b)   \; ({\bf 1} - \gamma^5) \gamma^\mu  \;
        \;
    v(p_d,\lambda_d)\; \;
        \bar u(p_c,\lambda_c)  \;  ({\bf 1} - \gamma^5) \gamma^\mu \;
         {-{\not\!k_2} \over -2k_2p_a} \not\!{\epsilon}^\star_{\sigma_2}(k_2)\;
    u(p_a,\lambda_a)  \Bigr)\\ &(p_d-p_b-p_c+p_a-k_2) \cdot
     {\epsilon}^\star_{\sigma_1}(k_1) \\
   &  + (1 \leftrightarrow 2),
  \end{split}
\end{equation}

and the second term 
\begin{equation}
  \label{isr-W-kc}
  \begin{split}
    L_{e^-,W}^{k_1,k_2}(2)= & (eQ_e)^2 BW_W\bigl((p_c+k_2-p_a)^2\bigr)  BW_W\bigl((p_c+k_2+k_1-p_a)^2\bigr)\\
\Bigl( 
    &   \bar v(p_b,\lambda_b)   \; ({\bf 1} - \gamma^5) \gamma^\mu  \;
        \;
    v(p_d,\lambda_d)\\ 
 &   2\bigl[  
    -\bigl( {\epsilon}^\star_{\sigma_1}(k_1) \bigr)_\nu (k_1)_\mu 
    +\bigl( {\epsilon}^\star_{\sigma_1}(k_1) \bigr)_\mu (k_1)_\nu 
    \bigr]  \\
    &    \bar u(p_c,\lambda_c)  \;  ({\bf 1} - \gamma^5) \gamma^\nu \;
         {-{\not\!k_2} \over -2k_2p_a} \not\!{\epsilon}^\star_{\sigma_2}(k_2)\;
    u(p_a,\lambda_a)  \Bigr)\\
   &  + (1 \leftrightarrow 2)
  \end{split}
\end{equation}
is gauge invariant by itself. The third one is explicitelly less divergent in collinear configuration:
\begin{equation}
  \label{isr-W-kd}
  \begin{split}
    L_{e^-,W}^{k_1,k_2}(\bar 3)= & -(eQ_e)^2 BW_W\bigl((p_c+k_2-p_a)^2\bigr)  BW_W\bigl((p_c+k_2+k_1-p_a)^2\bigr)\\
\Bigl( 
    &   \bar v(p_b,\lambda_b)   \; ({\bf 1} - \gamma^5) \not\!{\epsilon}^\star_{\sigma_1}(k_1) \;
        \;
    v(p_d,\lambda_d)\\ 
     &    \bar u(p_c,\lambda_c)  \;  ({\bf 1} - \gamma^5) {\not\!p_a} \;
         {-{\not\!k_2} \over -2k_2p_a} \not\!{\epsilon}^\star_{\sigma_2}(k_2)\;
    u(p_a,\lambda_a)  \Bigr)\\
   &  + (1 \leftrightarrow 2).
  \end{split}
\end{equation}
Let us now turn to the other part of the amplitude, and again present it  
in a form of the sum: 
\begin{equation}
  \label{isr-W-00}
  \begin{split}
    L_{e^-,W}^{k_1,k_2}(k^0)=L_{e^-,W}^{k_1,k_2}(\bar 4)+L_{e^-,W}^{k_1,k_2}(5)+L_{e^-,W}^{k_1,k_2}(\bar 6)
  \end{split}
\end{equation}
where first term reads
\begin{equation}
  \label{isr-W-0b}
  \begin{split}
    L_{e^-,W}^{k_1,k_2}(\bar 4)= & (eQ_e)^2 BW_W\bigl((p_c+k_2-p_a)^2\bigr)  BW_W\bigl((p_c+k_2+k_1-p_a)^2\bigr)\\
\Bigl( 
    &   \bar v(p_b,\lambda_b)   \; ({\bf 1} - \gamma^5) \gamma^\mu  \;
        \;
    v(p_d,\lambda_d)\; \;
        \bar u(p_c,\lambda_c)  \;  ({\bf 1} - \gamma^5) \gamma^\mu \;
         {\not\!{p_a}+m  \over -2k_2p_a} \not\!{\epsilon}^\star_{\sigma_2}(k_2)\;
    u(p_a,\lambda_a)  \Bigr)\\ &(p_d-p_b-p_c+p_a-k_2) \cdot
     {\epsilon}^\star_{\sigma_1}(k_1) \\
   &  + (1 \leftrightarrow 2),
  \end{split}
\end{equation}

the second one
\begin{equation}
  \label{isr-W-0c}
  \begin{split}
    L_{e^-,W}^{k_1,k_2}( 5)= & (eQ_e)^2 BW_W\bigl((p_c+k_2-p_a)^2\bigr)  BW_W\bigl((p_c+k_2+k_1-p_a)^2\bigr)\\
\Bigl( 
    &   \bar v(p_b,\lambda_b)   \; ({\bf 1} - \gamma^5) \gamma^\mu  \;
        \;
    v(p_d,\lambda_d)\\ 
 &   2\bigl[  
    -\bigl( {\epsilon}^\star_{\sigma_1}(k_1) \bigr)_\nu ((k_1)_\mu -(p_b)_\mu) 
    +\bigl( {\epsilon}^\star_{\sigma_1}(k_1) \bigr)_\mu (k_1)_\nu 
    \bigr]  \\
    &    \bar u(p_c,\lambda_c)  \;  ({\bf 1} - \gamma^5) \gamma^\nu \;
         {\not\!{p_a}+m  \over -2k_2p_a} \not\!{\epsilon}^\star_{\sigma_2}(k_2)\;
    u(p_a,\lambda_a)  \Bigr)\\
   &  + (1 \leftrightarrow 2),
  \end{split}
\end{equation}
and the third 
\begin{equation}
  \label{isr-W-0d}
  \begin{split}
    L_{e^-,W}^{k_1,k_2}(\bar 6)= & (eQ_e)^2 BW_W\bigl((p_c+k_2-p_a)^2\bigr)  BW_W\bigl((p_c+k_2+k_1-p_a)^2\bigr)\\
\Bigl( 
    &   \bar v(p_b,\lambda_b)   \; ({\bf 1} - \gamma^5) \not\!{\epsilon}^\star_{\sigma_1}(k_1) \;
        \;
    v(p_d,\lambda_d)\\ 
     &    \bar u(p_c,\lambda_c)  \;  ({\bf 1} - \gamma^5)( {\not\!k_2}-{\not\!p_a} )\;
         {\not\!{p_a}+m  \over -2k_2p_a} \not\!{\epsilon}^\star_{\sigma_2}(k_2)\;
    u(p_a,\lambda_a)  \Bigr)\\
   &  + (1 \leftrightarrow 2).
  \end{split}
\end{equation}
The last two terms  can be modified further, and some terms neglected. 
One can check that these  terms
contribute at the level of $\frac{m_e}{\sqrt{s}}$ only, and that is why,  we will exclude them
from explicit considerations.
After these simplifications, we finally obtain gauge invariant by itself
\begin{equation}
  \label{isr-W-0e}
  \begin{split}
    L_{e^-,W}^{k_1,k_2}(5)= & (eQ_e)^2 BW_W\bigl((p_c+k_2-p_a)^2\bigr)  BW_W\bigl((p_c+k_2+k_1-p_a)^2\bigr)\\
\Bigl( 
    &   \bar v(p_b,\lambda_b)   \; ({\bf 1} - \gamma^5) \gamma^\mu  \;
        \;
    v(p_d,\lambda_d)\\ 
 &   2\bigl[  
    -\bigl( {\epsilon}^\star_{\sigma_1}(k_1) \bigr)_\nu (k_1)_\mu  
    +\bigl( {\epsilon}^\star_{\sigma_1}(k_1) \bigr)_\mu (k_1)_\nu 
    \bigr]  \\
    &    \bar u(p_c,\lambda_c)  \;  ({\bf 1} - \gamma^5) \gamma^\nu \;
         {\not\!{p_a}+m  \over -2k_2p_a} \not\!{\epsilon}^\star_{\sigma_2}(k_2)\;
    u(p_a,\lambda_a)  \Bigr)\\
   &  + (1 \leftrightarrow 2).
  \end{split}
\end{equation}
and now explicitelly less divergent in collinear configuration term
\begin{equation}
  \label{isr-W-0f}
  \begin{split}
    L_{e^-,W}^{k_1,k_2}(\bar 6)= & (eQ_e)^2 BW_W\bigl((p_c+k_2-p_a)^2\bigr)  BW_W\bigl((p_c+k_2+k_1-p_a)^2\bigr)\\
\Bigl( 
    &   \bar v(p_b,\lambda_b)   \; ({\bf 1} - \gamma^5) \not\!{\epsilon}^\star_{\sigma_1}(k_1) \;
        \;
    v(p_d,\lambda_d)\\ 
     &    \bar u(p_c,\lambda_c)  \;  ({\bf 1} - \gamma^5) {\not\!k_2} \;
         {\not\!{p_a}+m  \over -2k_2p_a} \not\!{\epsilon}^\star_{\sigma_2}(k_2)\;
    u(p_a,\lambda_a)  \Bigr)\\
   &  + (1 \leftrightarrow 2).
  \end{split}
\end{equation}

Let us now turn to the diagrams of double emission from the $W$, see fig. \ref{Wemi}, right-hand side. The corresponding amplitude
can be written as:
\begin{equation}
  \label{W-W}
  \begin{split}
    L_{W,W}^{k_1,k_2}= & (eQ_e)^2 BW_W\bigl((p_c-p_a)^2\bigr)  BW_W\bigl((p_c+k_1-p_a)^2\bigr) \\
                          &        BW_W\bigl((p_c+k_1+k_2-p_a)^2\bigr)\\
\Bigl( 
    &   \bar v(p_b,\lambda_b)   \; ({\bf 1} - \gamma^5) \gamma^\mu \;
        \;
    v(p_d,\lambda_d)\; \; \;
        \bar u(p_c,\lambda_c)  \;  ({\bf 1} - \gamma^5) \gamma^\nu \; u(p_a,\lambda_a)  \Bigr) \\ 
& \bigl[g_{\sigma\nu}(p-q)_\rho +g_{\nu\rho}(q-k_1)_\sigma +g_{\sigma\rho}(k_1-p)_\nu  \bigr] \\
&\bigl[g^\sigma_{\mu}(p'-q')_{\rho'} +g^\sigma_{\rho'}(q'-k_2)_\mu +g_{\mu\rho'}(k_2-p')^\sigma  \bigr]\\
& ({\epsilon}^\star_{\sigma_1}(k_1))^\rho ({\epsilon}^\star_{\sigma_2}(k_2))^{\rho'}\\
   &  + (1 \leftrightarrow 2)
  \end{split}
\end{equation}
where $p=-q'=p_d-p_b+k_2=-(p_c-p_a+k_1)$, $q=p_c-p_a=-(p_d-p_b+k_1+k_2)$ and
$p'=p_d-p_b=-(p_c-p_a+k_1+k_2)$.

As usual, we will represent this expression in a form of the sum:
\begin{equation}
  \label{W-Wa}
  \begin{split}
    L_{W,W}^{k_1,k_2}=L_{W,W}^{k_1,k_2}(\bar 1)
+L_{W,W}^{k_1,k_2}(\bar 2)+L_{W,W}^{k_1,k_2}(\bar 3)
+L_{W,W}^{k_1,k_2}(\bar 4)+L_{W,W}^{k_1,k_2}(\bar 5)
+L_{W,W}^{k_1,k_2}(\bar 6).
  \end{split}
\end{equation}
The first term, proportional to  Born amplitude multiplied 
by factor depending on polarization of the two photons
takes the form:
\begin{equation}
  \label{W-Wb}
  \begin{split}
    L_{W,W}^{k_1,k_2}(\bar 1)= & (eQ_e)^2 BW_W\bigl((p_c-p_a)^2\bigr)  BW_W\bigl((p_c+k_1-p_a)^2\bigr) \\
                          &        BW_W\bigl((p_c+k_1+k_2-p_a)^2\bigr)\\
\Bigl( 
    &   \bar v(p_b,\lambda_b)   \; ({\bf 1} - \gamma^5) \gamma^\mu \;
        \;
    v(p_d,\lambda_d)\; \; \;
        \bar u(p_c,\lambda_c)  \;  ({\bf 1} - \gamma^5) \gamma_\mu \; u(p_a,\lambda_a)  \Bigr) \\ 
&      (p-q) \cdot {\epsilon}^\star_{\sigma_1}(k_1)   
 \;\; (p'-q') \cdot ({\epsilon}^\star_{\sigma_2}(k_2)) \\
   &  + (1 \leftrightarrow 2).
  \end{split}
\end{equation}

Terms where dependence on   polarization of only one photon factorizes out 
from the amplitude take a form

\begin{equation}
  \label{W-Wc}
  \begin{split}
    L_{W,W}^{k_1,k_2}(\bar 2)= & (eQ_e)^2 BW_W\bigl((p_c-p_a)^2\bigr)  BW_W\bigl((p_c+k_1-p_a)^2\bigr) \\
                          &        BW_W\bigl((p_c+k_1+k_2-p_a)^2\bigr)\\
\Bigl( 
    &   \bar v(p_b,\lambda_b)   \; ({\bf 1} - \gamma^5) \gamma^\mu \;
        \;
    v(p_d,\lambda_d)\; \; \;
        \bar u(p_c,\lambda_c)  \;  ({\bf 1} - \gamma^5) \gamma^\nu \; u(p_a,\lambda_a)  \Bigr) \\ 
& 2\bigl[-({\epsilon}^\star_{\sigma_1}(k_1))_\nu (k_1)_\mu \;\; +(k_1)_\nu ({\epsilon}^\star_{\sigma_1}(k_1))_\mu \bigr] \;  \\
&  (p'-q') \cdot ({\epsilon}^\star_{\sigma_2}(k_2))\\
   &  + (1 \leftrightarrow 2),
  \end{split}
\end{equation}
and
\begin{equation}
  \label{W-Wd}
  \begin{split}
    L_{W,W}^{k_1,k_2}(\bar 3)= & (eQ_e)^2 BW_W\bigl((p_c-p_a)^2\bigr)  BW_W\bigl((p_c+k_1-p_a)^2\bigr) \\
                          &        BW_W\bigl((p_c+k_1+k_2-p_a)^2\bigr)\\
\Bigl( 
    &   \bar v(p_b,\lambda_b)   \; ({\bf 1} - \gamma^5) \gamma^\mu \;
        \;
    v(p_d,\lambda_d)\; \; \;
        \bar u(p_c,\lambda_c)  \;  ({\bf 1} - \gamma^5) \gamma^\nu \; u(p_a,\lambda_a)  \Bigr) \\ 
& 2\bigl[-({\epsilon}^\star_{\sigma_2}(k_2))_\nu (k_2)_\mu \;\; +(k_2)_\nu ({\epsilon}^\star_{\sigma_2}(k_2))_\mu \bigr] \;  \\
&   (p-q) \cdot {\epsilon}^\star_{\sigma_1}(k_1)\\
   &  + (1 \leftrightarrow 2).
  \end{split}
\end{equation}
The two last terms (\ref{W-Wc},\ref{W-Wd})are partially gauge independent, respectively for 
the polarization vector of the first and second photon. 
The remaining fully gauge dependent parts of the amplitude   
read:

\begin{equation}
  \label{W-Wcp}
  \begin{split}
    L_{W,W}^{k_1,k_2}(\bar 4)= & -(eQ_e)^2 BW_W\bigl((p_c-p_a)^2\bigr)  BW_W\bigl((p_c+k_1-p_a)^2\bigr) \\
                          &        BW_W\bigl((p_c+k_1+k_2-p_a)^2\bigr)\\
\Bigl( 
    &   \bar v(p_b,\lambda_b)   \; ({\bf 1} - \gamma^5)  {\not\!k_2}\;
        \;
    v(p_d,\lambda_d)\; \; \;
        \bar u(p_c,\lambda_c)  \;  ({\bf 1} - \gamma^5) \not\!{\epsilon}^\star_{\sigma_1}(k_1) \; u(p_a,\lambda_a)  \Bigr) \\ 
&  (p'-q') \cdot ({\epsilon}^\star_{\sigma_2}(k_2))\\
   &  + (1 \leftrightarrow 2),
  \end{split}
\end{equation}
and
\begin{equation}
  \label{W-Wdp}
  \begin{split}
    L_{W,W}^{k_1,k_2}(\bar 5)= & (eQ_e)^2 BW_W\bigl((p_c-p_a)^2\bigr)  BW_W\bigl((p_c+k_1-p_a)^2\bigr) \\
                          &        BW_W\bigl((p_c+k_1+k_2-p_a)^2\bigr)\\
\Bigl( 
    &   \bar v(p_b,\lambda_b)   \; ({\bf 1} - \gamma^5)  \not\!{\epsilon}^\star_{\sigma_2}(k_2)\;
        \;
    v(p_d,\lambda_d)\; \; \;
        \bar u(p_c,\lambda_c)  \;  ({\bf 1} - \gamma^5)  {\not\!k_1} \; u(p_a,\lambda_a)  \Bigr) \\ 
&   (p-q) \cdot {\epsilon}^\star_{\sigma_1}(k_1)\\
   &  + (1 \leftrightarrow 2).
  \end{split}
\end{equation}
Finally
\begin{equation}
  \label{W-We}
  \begin{split}
    L_{W,W}^{k_1,k_2}(\bar 6)= & (eQ_e)^2 BW_W\bigl((p_c-p_a)^2\bigr)  BW_W\bigl((p_c+k_1-p_a)^2\bigr) \\
                          &        BW_W\bigl((p_c+k_1+k_2-p_a)^2\bigr)\\
\Bigl( 
    &   \bar v(p_b,\lambda_b)   \; ({\bf 1} - \gamma^5) \gamma^\mu \;
        \;
    v(p_d,\lambda_d)\; \; \;
        \bar u(p_c,\lambda_c)  \;  ({\bf 1} - \gamma^5) \gamma^\nu \; u(p_a,\lambda_a)  \Bigr) \\ 
& \bigl[g_{\nu\rho}(q-k_1)_\sigma +g_{\sigma\rho}(k_1-p)_\nu  \bigr] \; ({\epsilon}^\star_{\sigma_1}(k_1))^\rho \\
&\bigl[g^\sigma_{\rho'}(q'-k_2)_\mu +g_{\mu\rho'}(k_2-p')^\sigma  \bigr]\;  ({\epsilon}^\star_{\sigma_2}(k_2))^{\rho'}\\
   &  + (1 \leftrightarrow 2).
  \end{split}
\end{equation}


At the last step, let us turn to contributions from diagrams presented in 
fig.~\ref{Hemi}. The diagram with  contribution from quatric gauge coupling reads:
\begin{equation}
  \label{W-qatr}
  \begin{split}
    L_{W^2}^{k_1,k_2}= & (eQ_e)^2 BW_W\bigl((p_c-p_a)^2\bigr)  BW_W\bigl((p_c+k_2+k_1-p_a)^2\bigr)\\
   \Bigl(  
    &   \bar v(p_b,\lambda_b)   \; ({\bf 1} - \gamma^5) \not\!{\epsilon}^\star_{\sigma_1}(k_1) \;
        \;
    v(p_d,\lambda_d)
     \;\;    \bar u(p_c,\lambda_c)  \;  ({\bf 1} - \gamma^5)  \;
          \not\!{\epsilon}^\star_{\sigma_2}(k_2)\;
    u(p_a,\lambda_a)   \\
    +&   \bar v(p_b,\lambda_b)   \; ({\bf 1} - \gamma^5) \not\!{\epsilon}^\star_{\sigma_2}(k_2) \;
        \;
    v(p_d,\lambda_d)
     \;\;    \bar u(p_c,\lambda_c)  \;  ({\bf 1} - \gamma^5)  \;
          \not\!{\epsilon}^\star_{\sigma_1}(k_1)\;
    u(p_a,\lambda_a)   \\
   + &  \bar v(p_b,\lambda_b)   \; ({\bf 1} - \gamma^5) \gamma^\mu \;
        \;
    v(p_d,\lambda_d)
     \;\;    \bar u(p_c,\lambda_c)  \;  ({\bf 1} - \gamma^5)  \;
          \gamma_\mu\;
    u(p_a,\lambda_a)  
   {\epsilon}^\star_{\sigma_1}(k_1) \cdot {\epsilon}^\star_{\sigma_2}(k_2) \Bigr).
  \end{split}
\end{equation}

It is convenient to write it as a sum of two parts
\begin{equation}
  \label{W-qatra}
  \begin{split}
  L_{W^2}^{k_1,k_2}=  L_{W^2}^{k_1,k_2}(\bar 1)+  L_{W^2}^{k_1,k_2}(\bar 2)
 \end{split}
\end{equation}
where
\begin{equation}
  \label{W-qatrb}
  \begin{split}
    L_{W^2}^{k_1,k_2}(\bar 1)= & 2(eQ_e)^2 BW_W\bigl((p_c-p_a)^2\bigr)  BW_W\bigl((p_c+k_2+k_1-p_a)^2\bigr) \;\;
      {\epsilon}^\star_{\sigma_1}(k_1) \cdot {\epsilon}^\star_{\sigma_2}(k_2) \\
    &  \bar v(p_b,\lambda_b)   \; ({\bf 1} - \gamma^5) \gamma^\mu \;
        \;
    v(p_d,\lambda_d)
     \;\;    \bar u(p_c,\lambda_c)  \;  ({\bf 1} - \gamma^5)  \;
          \gamma_\mu\;
    u(p_a,\lambda_a)   
  \end{split}
\end{equation}

and
\begin{equation}
  \label{W-qatrc}
  \begin{split}
    L_{W^2}^{k_1,k_2}(\bar 2)= & (eQ_e)^2 BW_W\bigl((p_c-p_a)^2\bigr)  BW_W\bigl((p_c+k_2+k_1-p_a)^2\bigr)\\
   \Bigl(  
    &   \bar v(p_b,\lambda_b)   \; ({\bf 1} - \gamma^5) \not\!{\epsilon}^\star_{\sigma_1}(k_1) \;
        \;
    v(p_d,\lambda_d)
     \;\;    \bar u(p_c,\lambda_c)  \;  ({\bf 1} - \gamma^5)  \;
          \not\!{\epsilon}^\star_{\sigma_2}(k_2)\;
    u(p_a,\lambda_a)   \\
    +&   \bar v(p_b,\lambda_b)   \; ({\bf 1} - \gamma^5) \not\!{\epsilon}^\star_{\sigma_2}(k_2) \;
        \;
    v(p_d,\lambda_d)
     \;\;    \bar u(p_c,\lambda_c)  \;  ({\bf 1} - \gamma^5)  \;
          \not\!{\epsilon}^\star_{\sigma_1}(k_1)\;
    u(p_a,\lambda_a)  \Bigr).
  \end{split}
\end{equation}

Contribution from the  diagram involving internal $\chi$  line reads:
\begin{equation}
  \label{W-H}
  \begin{split}
    L_{W,\chi}^{k_1,k_2}= & (eQ_e)^2 M_W^2\; BW_W\bigl((p_c-p_a)^2\bigr)  BW_W\bigl((p_c+k_2+k_1-p_a)^2\bigr)\\
\Bigl( 
    &   \bar v(p_b,\lambda_b)   \; ({\bf 1} - \gamma^5) \not\!{\epsilon}^\star_{\sigma_1}(k_1) \;
        \;
    v(p_d,\lambda_d)\\ 
     &    \bar u(p_c,\lambda_c)  \;  ({\bf 1} - \gamma^5)  \;
          \not\!{\epsilon}^\star_{\sigma_2}(k_2)\;
    u(p_a,\lambda_a)  \Bigr)\\
   &  + (1 \leftrightarrow 2).
  \end{split}
\end{equation}

This closes the list of all diagrams entering the complete spin amplitude for the process 
$e^+e^- \to \nu_e \bar \nu_e \gamma \gamma$. The contributing terms were obtained 
from the Feynman rules, and 
 were grouped on the basis of rather straightforward rules; gauge symmetry and nature
of singularities in infrared and collinear limits (phase space integration was not necessary).

The complete gauge invariant part of the spin amplitude of $W$ exchange  can be now written as:
\begin{equation}
 \label{Wgg-all}
  \begin{split}
\cal{M}_W=\cal{M}_W^A+\cal{M}_W^B,
  \end{split}
\end{equation}
where $\cal{M}_W^A$ (technically  identical to the amplitude of $Z$ exchange)  
was given by formula (\ref{Wgg-a}), and new part, specific to the $W$ bosonic interactions, reads:

\begin{equation}
 \label{Wgg-all-B}
  \begin{split}
{\cal M}_W^B=  
& L_{e^+,W}^{k_1,k_2}(\bar 1)+L_{e^+,W}^{k_1,k_2}(2)+L_{e^+,W}^{k_1,k_2}(\bar 3)+
  L_{e^+,W}^{k_1,k_2}(\bar 4)+L_{e^+,W}^{k_1,k_2}(5)+L_{e^+,W}^{k_1,k_2}(\bar 6)+ \\
& L_{e^-,W}^{k_1,k_2}(\bar 1)+L_{e^-,W}^{k_1,k_2}(2)+L_{e^-,W}^{k_1,k_2}(\bar 3)+
  L_{e^-,W}^{k_1,k_2}(\bar 4)+L_{e^-,W}^{k_1,k_2}(5)+L_{e^-,W}^{k_1,k_2}(\bar 6)+ \\
& L_{W,W}^{k_1,k_2}(\bar 1)+L_{W,W}^{k_1,k_2}(\bar 2)+L_{W,W}^{k_1,k_2}(\bar 3)+L_{W,W}^{k_1,k_2}(\bar 4)+L_{W,W}^{k_1,k_2}(\bar 5)+L_{W,W}^{k_1,k_2}(\bar 6)+ \\
& L_{W^2}^{k_1,k_2}(\bar 1)+L_{W^2}^{k_1,k_2}(\bar 2)+L_{W,\chi}^{k_1,k_2}.
  \end{split}
\end{equation}

We can now write the complete spin amplitude, of  $W$ interactions, 
as a sum of gauge invariant parts:
\begin{equation}
 \label{Wgg-all-sum}
  \begin{split}
\cal{M}_W=& {\cal M}_1+ {\cal M}_2+ {\cal M}_3+  {\cal M}_4+ {\cal M}_5+ {\cal M}_6+ {\cal M}_7 +\\
       & {\cal M}_8+{\cal M}_9 +{\cal M}_{10}+{\cal M}_{11},
  \end{split}
\end{equation}

where
\begin{equation}
 \label{Wgg-all-sum-a}
  \begin{split}
{\cal M}_1=& L_{e^-}^{k_1,k_2}(1)\\
{\cal M}_2=& L_{e^-}^{k_1,k_2}(2)\\
{\cal M}_3=& L_{e^+}^{k_1,k_2}(1)\\
{\cal M}_4=& L_{e^+}^{k_1,k_2}(2)\\
{\cal M}_5=& L_{e^-,e^+}^{k_1,k_2}(1)\\
{\cal M}_6=& L_{e^-}^{k_1,k_2}(\bar 4)+L_{e^+}^{k_1,k_2}(\bar 4)+L_{e^-,e^+}^{k_1,k_2}(\bar 2)+ L_{e^-,W}^{k_1,k_2}(\bar 1)+ L_{e^+,W}^{k_1,k_2}(\bar 1) \\
{\cal M}_7=& L_{e^-}^{k_1,k_2}(\bar 3)+L_{e^+}^{k_1,k_2}(\bar 3)+L_{e^-,e^+}^{k_1,k_2}(\bar 3) +L_{e^-,W}^{k_1,k_2}(\bar 4)+L_{e^+,W}^{k_1,k_2}(\bar 4)+L_{W,W}^{k_1,k_2}(\bar 1)+L_{W^2}^{k_1,k_2}(\bar 1)\\
{\cal M}_8=& L_{e^-,W}^{k_1,k_2}(2)\\
{\cal M}_9=& L_{e^+,W}^{k_1,k_2}(2) \\
{\cal M}_{10}=& L_{e^-,W}^{k_1,k_2}(5)+L_{e^+,W}^{k_1,k_2}(5)+L_{W,W}^{k_1,k_2}(\bar 2)+L_{W,W}^{k_1,k_2}(\bar 3)\\
{\cal M}_{11}=&L_{e^-,W}^{k_1,k_2}(\bar 3)+L_{e^+,W}^{k_1,k_2}(\bar 3)+L_{e^-,W}^{k_1,k_2}(\bar 6)+ L_{e^+,W}^{k_1,k_2}(\bar 6)
+L_{W,W}^{k_1,k_2}(\bar 4)+L_{W,W}^{k_1,k_2}(\bar 5)+L_{W,W}^{k_1,k_2}(\bar 6)+\\
&L_{W^2}^{k_1,k_2}(\bar 2)+L_{W,\chi}^{k_1,k_2}.
  \end{split}
\end{equation}

The matching of the $L^a_b(\bar n)$ terms into gauge invariant parts ${\cal M}_i$ of the
amplitude is straigtforward and based on the type of singularities present/absent  
in the particular group.
Each of the listed below contributions ${\cal M}_1$--${\cal M}_{11}$ can be given 
some physical interpretation. In some cases, appearance of such parts may seem 
rather unexpected. In brackets 
we provide symbols such as {\tt (IA)}, they  denote the name of variables 
 used in {\tt KKMC} \cite{Jadach:1999vf} Monte Carlo,  as keys for the parts of the amplitude:
\begin{itemize}
\item
${\cal M}_1$ {\tt (IA)},
contribution of the infrared non-singular contributions of double emission from
electron line, part with straightforward gauge cancellation within the terms originating
from diagram of two photons attached to the same incoming electron line.
\item
${\cal M}_2$ {\tt (IV2)}
contribution of the infrared non-singular contributions of double emission from
electron line, part with non-straightforward gauge cancellation within the terms originating
from diagram of two photons attached to the same incoming electron line. 
Part of the diagram contribution had to be subtracted; more precisely
expression without  $k_1k_2$ product
in electron propagator. This subtraction term is recupered 
in ${\cal M}_6$ and ${\cal M}_7$. 
\item
${\cal M}_3$ {\tt (IA)},
${\cal M}_4$ {\tt (IV1)}
as in previous two cases but for emission from positron line.
\item 
${\cal M}_5$  {\tt (I8)} infrared non-singular contributions of single emission from
 electron- and another single emission from positron line. This contribution is gauge invariant by construction.
\item 
${\cal M}_6$ {\tt (I9X), (I9Y), (I9Z), (I9T)}, part of the amplitude with  infrared factor for
one photon, and for the second one infrared non-singular gauge invariant
contribution. For the diagrams with $W$ exchange contribution from diagram with 
photon emission from $W$ need to be taken. For the gauge cancellation to hold, relation between
$t$-channel transfers in $W$ propagators and momenta multiplying photon polarization vector need to be fulfiled.
Nonetheless certain freedom in choice is left. It was useful in construction of extrapolation procedures\footnote{ 
Identical  condition, also originating directly from Ward-identities, need to be preserved 
in ${\cal M}_{10}$ and the similar one in ${\cal M}_{11}$.}.
\item
${\cal M}_7$ {\tt (IVI)}
part of the amplitude with  infrared factors for
both photons.  For the diagrams with $W$ exchange contribution from diagrams with 
single and double emission of photons  from $W$ needs to be taken, also part of the diagram with
quartic gauge coupling was  needed here. 
\item
${\cal M}_8$ {\tt   (I71), (I72)} part of the amplitude with infrared non-singular contribution of emission from electron
for one photon and for  
another one part of emission from $W$ which is self gauge-conserving.
\item
${\cal M}_9$ {\tt  (I71), (I72)}
as in previous case but for emission from positron.
\item
${\cal M}_{10}$ {\tt (I9s1), (I9s2) } part of the amplitude with  infrared factor for
one photon and for
 another one  part of emission from $W$ which is self gauge-conserving
part.
\item
${\cal M}_{11}$ {\tt (I9), (I9B), (I10)} all remaining parts, they turn out to be  
free of singularities both in collinear and soft limits.
\item
Let us comment that in the limit $M_W \to \infty$ all contributions from ${\cal M}_6$ to ${\cal M}_{11}$ 
disappear.
In this limit
amplitudes for $s$-channel $Z$ exchange and $t$-channel $W$ nearly coincide. The only remaining difference 
is the coupling constants and hard interaction part of the amplitude given respectively 
by  formulas (\ref{cdotsZ}) and  (\ref{cdotsW}).
This is an extension of similar observation of reference \cite{Richter-Was:1994ta} instrumental in construction 
of extrapolation procedures of ref.~\cite{Jadach:1998wp}, to the case beyond real photon interactions with fermions only.
\item
Let us point that in many places we have used separation of the $WW\gamma$ vertex into three parts;
(i) the one with the $g_{\mu\nu}$ tensor along line connecting fermion lines, (ii) the part internally
preserving gauge symmetry, (iii) the remaining part which we often could reduce significantly with the help of Dirac
equation (for the fermion lines connected with the $WW\gamma$ vertex by $W$ propagator.
\item
Finally let us note, that the above separation into gauge invariant parts can be continued even further. 
For example it is rather easy to separate ${\cal M}_6$ into four parts. 
For each,  emissions of 
individual photons are attributed {\it either} to electron {\it or} positron line.

\end{itemize}

Let us note that we have not exploited to the end the properties of ${\cal M}_{11}$. It was not interesting 
from the point of view of our main purpose, which is implementation of the matrix element to the environment of 
Coherent Exlusive Exponentiation. Also in case of ${\cal M}_{11}$, contrary to the cases  ${\cal M}_{1 }$ to ${\cal M}_{10}$,
similarities with first order results could not be seen. This is rather natural, as for example 
quatric gauge couplings are absent in first order. In this case  
hint on pattern of  constructing amplitudes
of even higher order using iteration techniques could not be found. To this end, discussion of the 
amplitudes of triple photon emission would be needed. If conclusive, it would point to solutions beyond 
next-to-leading-log approximation, thus beyond imminent interest of the present paper.

The gauge invariance was not the only element of the criterium which was used here to split amplitude into gauge invariant parts. Equally important was that the
two main sources of the radiation, incoming beams, form the unambiguous 
frame with respect to which, photon energy  and  the 
angles of photons with respect to fermions could be defined. That is why there was no need 
to make any reference to the regulators, singular terms could be localized
already at the amplitude level and in fully differential manner, with no need
to partially integrate phase space.  The expansion in the contact 
interaction for $W$ propagator enabled to place the gauge cancellation
effects of emission from $t$-channel $W$ within the frame of ISR radiation.
Also relation between amplitude for double and single photon emission had
to be exploited to close down window for ambiguities.
Once these assumptions and properties were exploited, the solution  seem 
to be unique, up to may be grouping or further splitting of the obtained parts. 
Confirmation, whether this is accidental property and observation
which hold for the particular cases and up to the second order only, 
may require calculation to be extended to at least third order.

\section{Some points on extrapolation }
Let us summarize here some  specific issues related to extrapolation procedure 
of CEEX scheme described in detail in ref.~\cite{Jadach:2000ir} for purely $s$-channel 
hard process.
 One of the important 
property of perturbation expansion,   rearranged 
to improve convergence into exclusive exponentiation, is that parts 
of the amplitudes  need 
to be appropriately shifted between the rearranged orders of expansion. 
We will concentrate on issues related
to real bremsstrahlung only.
In particular parts of the higher order terms 
(directly calculated in a standard way) 
which are already available at lower level of CEEX perturbation expansion  
need to be localized and subtracted in a clear 
way. Only remaining residual parts, called $\beta^0$, $\beta^1$,   $\beta^2$ etc. 
\cite{yfs:1961} will be indeed the term of the given newly rearranged
order. The use of   $\beta$ functions is unambiguous if sufficiently 
high order of standard perturbative calculation is available to calculate
matrix element, for the configuration with all real photons. 
However it is not always the case, practical solutions for  exponentiation 
require definition of
methods how to calculate
matrix elements for the kinematical configuration with large number of real 
photons, using results of first (or second) order of perturbation
expansion only. 

There are several rules which extrapolation procedure must fulfill. Already
the lowest order must include all terms with the highest power of infrared 
singularity and for all kinematical configurations of arbitrary 
number of real photons and in a fully exclusive manner.  
Then, first order provide all terms
with next to highest power of infrared singularity, etc. Let us stress that 
reduction/extrapolation  procedure of exponentiation offers some freedom
of choice. This freedom can be used to  further improve convergence 
of perturbation expansion. The best guidance is of course comparison with 
result of even higher order of expantion to minimize their contribution.
 If such results are not available, 
higher order leading log results can be used instead. Finally, let us stress
that if sufficiently high order of perturbation expansion is available, 
dependence on particular choice of extrapolation drops out and unique result,
identical to the one of direct perturbation expansion without any reordering,
will be obtained.
Unfortunately this is not expected to be the case in  foreseeable future.

In case of diagrams with $Z$ exchange the question of choice of
extrapolation  procedure is straightforward. Inspection
of first order (formula \ref{isr-feynman1}) and second order (formula  \ref{Zgg-b}) amplitudes 
points to the following solution: the terms ${\cal M}_1$ to ${\cal M}_5$
of (\ref{Zgg-b}) should contribute to $\beta^2$, whereas the last two terms
${\cal M}_6$ and ${\cal M}_7$ can be directly obtained from the lower order.
The ${\cal M}_6$ can be obtained from 
 $\beta^1$  by multiplication with  the  soft photon factor for
the other photon. The
 $\beta^1$ can be identified as 
 this part of ${\cal M}^0$  (see formula \ref{isr-feynman1}) 
which is proportional to $\not\!k_1$. 
The  ${\cal M}_7$ can be obtained from the lowest order Born spin amplitude $\beta^0$
of $Z$ exchange  by  multiplication with 
two soft photon factors, for each of the bremsstrahlung photons,
exactly as it should be in exponentiation prescription.
The factorization properties
 can be easily seen if rather trivial manipulation on Dirac algebra is 
performed.

In case of diagrams with $W$ exchange, the question of choice of extrapolation 
procedure is slightly more complex, because of  dependence 
of the transfers  used in $W$ propagators on photon momenta. 
That is also the reason why triple and quatric gauge couplings need to be included 
in the considerations. If the kinematical configurations of more than two explicit 
hard photons are taken, then the transfers calculated for $W$ propagators can be defined 
in several ways.
Our choice used at present in KKMC \cite{Jadach:1999vf} is inspired by
 leading log considerations. For  lowest order ($\beta^0$,) and 
if there was no addition photons,  transfer $t_0$,
can be calculated  either as (i) $t_0=(p_c-p_a)^2$ 
or  (ii) $t_0=(p_d-p_b)^2$. If there is a photon collinear to $p_b$ 
the first choice is closer to the transfer dominating higher order 
(i.e. single bremsstrahlung) spin amplitude. In general the choice (i)
is thus more favored if total four-momentum carried out by the
sum of all photons is
pointing rather into direction of $p_b$  than $p_a$. Otherwise
the second choice is better. In case of single (or double) photon emission
 the  choice  how 
the transfers are calculated is basically the same. 
The only difference is, that the photons explicitely included in the particular contribution 
to $\beta^1$ or $\beta^2$, should contribute to the sum of photons mentioned above.
The choice which pair of 
four momenta  ($p_a$, $p_c$ or  $p_b$, $p_d$) is used in calculation for transfers,
 must  be taken in calculation 
of algebraic expressions originating from direct $W$ interaction with photons,
for  gauge invariance  to  hold.

\section{Summary}

We have presented complete results for the spin amplitudes 
of  $e^+ e^- \to \nu_e \bar \nu_e \gamma \gamma$ process. 
Using gauge transformation, as well as expansion with respect to contact
approximation for $W$ exchange, we were able to identify gauge invariant parts 
of amplitudes 
of the well defined physical properties. In particular the
parts proportional to inverse of photon energies 
(i.e. corresponding to infrared singularity), remaining parts proportional
to inverse of the product of fermion and photon momenta (i.e. of the type of collinear 
singularity) as well as residual finite parts could be grouped together in a rather natural 
 way.
By comparison with amplitudes for the diagrams involving $s$-channel $Z$ exchange we were
able to observe certain pattern of universality for many of those terms.  

Let us stress, that
some of the results presented here, could  be expected from the properties of U(1) 
gauge symmetry and the corresponding Ward identities. They are known already since 
a long time, also in the context of QCD.  The purpose of the present paper 
is mainly technical.
 We illustrate the scheme of step-by-step gauge cancellations and how they work 
for spin amplitude techniques.  
 Finally, we show, how they helped to develop
extrapolation procedures used in the KK Monte Carlo in case of neutrino channel.

Let us point to the  similar observation \cite{shimizu} as ours,
   in case of  the single loop corrections,
again for 
 $e^+e^- \to \nu_e\bar\nu_e$ process  and  also for $e^+e^- \to \nu_e\bar\nu_eH$.

\section*{Acknowledgments}
I would like to thank 
W. P\l{}aczek and Y. Shimizu, 
for useful discussions, and Y. Kurihara, for help in numerical comparisons with Grace
package.



\newpage
\providecommand{\href}[2]{#2}


\end{document}